\renewcommand\b{\beta}

\newcommand{\diracslash}[1]{#1\llap{/\kern2pt}}

\newcommand{\be}{\begin{equation}}
\newcommand{\ee}{\end{equation}}
\newcommand{\bea}{\begin{eqnarray}}
\newcommand{\eea}{\end{eqnarray}}
\newcommand{\ba}[1]{\begin{array}{#1}}
\newcommand{\ea}{\end{array}}

\newcommand{\bt}{\begin{tabular}}
\newcommand{\et}{\end{tabular}}

\newcommand{\beas}{\begin{eqnarray*}}
\newcommand{\eeas}{\end{eqnarray*}}

\documentclass[amsmath,12pt,amssymb,preprint,nofootinbib]{revtex4}
\usepackage{amsfonts} 
\usepackage{graphicx} 
\usepackage{epsfig}
\usepackage{multirow}
\usepackage{bm}

\begin{document}

\title{\textbf{Medium modifications of Heavy Quarkonia masses\\
in a generalized Linear Sigma Model}}
\author{Arpita Mondal}
\email{arpita.mondal1996@gmail.com}
\author{Pallabi Parui}
\email{pallabiparui123@gmail.com}
\author{Amruta Mishra}
\email{amruta@physics.iitd.ac.in}  
\affiliation{Department of Physics, 
Indian Institute of Technology, Delhi, Hauz Khas, New Delhi - 110016}
\begin{abstract}
We study the mass shifts of the charmonium ($\bar{c}c$) states 
($J/\psi$, $\psi(2S)$, $\psi(1D)$, $\chi_{c0}$, $\chi_{c1}$ and $\chi_{c2}$)
as well as the bottomonium ($\bar{b}b$) states ($\Upsilon(1S)$, 
$\Upsilon(2S)$, $\Upsilon_2(1D)$, $\chi_{b0}$, $\chi_{b1}$ 
and $\chi_{b2}$) in isospin asymmetric nuclear matter. These are
investigated using a generalized linear sigma model. 
The broken scale invariance of QCD is incorporated in the 
chiral $SU(2)\times SU(2)$ Lagrangian 
through an effective potential involving logarithmic terms 
of a scalar (glueball) dilaton field $\chi$. The mass shifts of the 
quarkonium states are obtained through 
the medium modifications of the dilaton field 
which simulates the scalar gluon condensate of QCD. 
We observe an appreciable mass drop in the states of heavy quarkonia under this study. The in-medium masses at finite densities thus obtained should modify the in-medium partial decay widths of heavy quarkonia to open heavy flavor 
mesons. These density effects can be probed in 
in the high energy nuclear collisions 
at the future facility at GSI (at Germany) and JINR (at Russia)
in the experiments producing highly dense baryonic matter.
 
\end{abstract}
\maketitle

\section{Introduction}
\label{I}
The study of the in-medium properties of heavy flavor mesons is a very consequential topic of research in strong interaction physics.
Such medium modifications of the properties like masses and decay widths of the produced heavy flavor mesons can further affect the experimental observables, such as production and propagation of the particles through the medium, created in the heavy ion collision (HIC) experiments at various high energy particle accelerators \cite{Hosaka_Prog_Part_Nucl_Phys}. In the HIC experiments, a highly dense medium is generated due to the collision of two relativistic, high energetic heavy ion beams.
As these experiments involve nuclei with large difference in the numbers of neutrons and protons, isospin asymmetry is an important parameter to be included in the study.
All of these in-medium hadronic properties can be well understood theoretically by the structure of QCD vacuum \cite{vacuum}, which is assumed to be densely populated by the chargeless and colorless states of strongly bounded light quark-antiquark pairs and gluons due to the spontaneous chiral symmetry breaking and scale invariance breaking in QCD vacuum. Their amplitudes of fluctuation so large that they can not be described perturbatively. Thus the corresponding vacuum is called non-perturbative vacuum and the strength of their fluctuations in vacuum 
are represented by vacuum expectation values (or vacuum condensates), 
as long as the distance between the quark and anti-quark is assumed 
to be small as compared to the scale of medium fluctuations. 

In the present work we investigate the medium modifications of the masses of the heavy quarkonia (the bound states of the heavy quark (c or b) and its antiquark partner ($\bar{c}$ or $\bar{b}$)), i.e., charmonia and bottomonia within the asymmetric nuclear matter. The medium modifications of the masses of open heavy flavor mesons ($D,~B$) arise due to the medium modifications of the nonzero light quark condensates, through the light quark (anti-quark) constituents of these mesons. For the heavy quarkonium states, the mass modifications are due to the gluon condensates \cite{pes1,pes2}. Theoretically, the heavy quark condensates are expressed entirely by the gluonic operators, in the heavy quark systems, through the operator product expansion (OPE) of the correlation function of the heavy quark currents, in the massless quark limit.  
Thus the in-medium leading order mass shift, in the zero quark-mass limit, is dependent on the difference of the scalar gluon condensate in the nuclear matter from its vacuum value \cite{leeko}. 
The in-medium gluon condensates are calculated in the current study within a chiral $SU(2)\times SU(2)$ Lagrangian, which incorporates the QCD trace anomaly through an effective potential involving logarithmic terms by introducing the scalar glueball field $\chi$.
This approach is a generalized version of the linear $\sigma$ model to account for the QCD trace anomaly effect, as it is depicted in refs. \cite{ellis, Heide, heide1992}, in the study of nuclear matter and finite nuclei under the mean field approximation of meson fields. The effects of baryon density and isospin asymmetry are incorporated through the number density, scalar density and isospin number density of nucleons in the scalar fields equations of motion under this mean field approximation. As a result of these mass modifications, some decay channels which are not possible in vacuum, can show up and some decay channels which are obvious in vacuum, can stop in the highly dense medium by the energy-momentum conservation law. This affects the partial decay widths, decay modes and the total decay width of the corresponding state. Moreover, it modifies the production yield of the charm and bottom states in the experiments. If the states are decaying into leptonic/semileptonic states then these medium modifications will be reflected on the resulted dilepton spectra and leptonic events. Furthermore, due to the nature of the interactions of the states in the medium there is also a possibility to observe the nuclear bound quarkonium states \cite{Tsushima}. 

Experimentally, the heavy quarkonium states are produced in the beginning of the ultra relativistic heavy ion collision (HIC) experiments such as, Relativistic Heavy Ion Collider (RHIC) at  Brookhaven National Laboratory (BNL) and Large Hadron Collider (LHC) at European Organization for Nuclear Research (CERN), where in the central (or near central) collisions, the highly dense medium is produced and due to the non-central collisions, a relatively low dense medium with the very large magnetic field is estimated to be produced in the perpendicular direction of the reaction plane depending upon the impact parameter, kind of heavy ions used and the energy of the collision \cite{HIC}. The upcoming accelerator FAIR (Facility for Antiproton and Ion Research) at GSI \cite{GSI1} in Germany will study the heavy mesons under the dense nuclear medium. Mainly, the CBM (Compressed Baryonic Matter) experiments intends to study the near threshold production of D-mesons and charmonium states in HIC experiments. Due to the energy range of CBM experiment, the underlying physics of bottomonia and open bottom mesons can not be explored there. Wherein, PANDA (anti-Proton ANihilation at DArmstadt) Collaboration focuses on heavy meson spectroscopy as well as on production by anti-proton annihilation in nuclei \cite{GSI} and as a high energy storage ring collider, it is feasible to produce the bottom mesons \cite{frankfurt} in this experiment. The CEBAF (Continuous Electron Beam Acclerator Facility) accelerator at JLab (Jefferson Laboratory) \cite{jlab} also studies the production of charm hadrons. In the future experiments at the NICA \cite{nica} (Nuclotron-based Ion Collider Facility) at JINR (Joint Institute of Nuclear Research), Russia, the hadronic matter is planned to be produced at large densities and at its energy range the bottomonia and open bottom meson are very likely to be produced \cite{nica1,nica2}.

The wave functions of the charmonium and the bottomonium states can be obtained mathematically by solving the Schr\"odinger equation using the harmonic oscillator potential \cite{arfken}. Our current study involves the mass shifts of the $S$-wave ($J/\psi$, $\psi(2S)$), P-wave ($\chi_{c0}$, $\chi_{c1}$, $\chi_{c2}$) and D-wave ($\psi(1D)$) states of charmonium and the S-wave ($\Upsilon(1S)$, $\Upsilon(2S)$), P-wave ($\chi_{b0}$, $\chi_{b1}$, $\chi_{b2}$) and D-wave ($\Upsilon_2(1D)$) (in the latest Particle Data Group review \cite{pdg}, this state is mentioned as the former $\Upsilon(1D)$ state with same mass) states of bottomonium in highly dense symmetric and asymmetric nuclear matter. The in-medium values of the scalar glueball field $\chi$, from which the modifications of gluon condensate come through, are studied within the framework of an effective chiral Lagrangian \cite{heide1992, Heide}.  
At the leading order, the mass shifts of charmonia and bottomonia are obtained 
from the medium modifications of the $\chi$ field. 
In reference \cite{leeko}, a QCD second order stark effect approach has been taken to describe the mass shifts of the charmonium states, $J/\psi$, $\psi(2S)$ and $\psi(1D)$.
The in-medium properties of the heavy flavor mesons have been studied in the hadronic medium using the QCD Sum Rules (QSR) approach
\cite{open_heavy_flavour_qsr,kimlee,klingl}. A combined approach 
of chiral SU(3) model and QCD sum rule method shows 
significant modifications for the charmonium properties 
\cite{amarvjpsi_qsr}. The in-medium properties of charmonium \cite{betachi,amarvdmesonTprc, amarvepja}, bottomonium \cite{DAM1}, open bottom \cite{DAM2} and open charm \cite{1,AMC} mesons are also studied using a chiral effective model.
Using the potential model 
\cite{eichten_1,eichten_2,Klumberg_Satz_Charmonium_prod_review,Quarkonia_QGP_Mocsy_IJMPA28_2013_review,repko} heavy quarkonium states are also investigated.
In the Quark Meson Coupling (QMC) model
\cite{QMC_Krein_Thomas_Tsushima_Prog_Part_Nucl_Phys_2018, QMC1, QMC2} the interactions of the mesons in the medium is considered through the couplings with quarks and the study also shows a mass drop of the mesons as well as baryons with density in the medium. In \cite{F1, F2}, the partial decay widths of heavy quarkonium states are studied within a hadronic medium, using the field-theoretic model of composite hadrons with quark (and antiquark) constituents. Furthermore, in the coupled channel approach \cite{tolos_heavy_mesons}, the resonances and the interaction mechanisms are studied. 
The chiral effective model has been used to several studies like, in finite nuclei \cite{2}, the in-medium properties of light vector mesons \cite{3,APS}, kaon, anti-kaon in the interior of (proto-) neutron stars \cite{4}.

The paper is organized as: In section \ref{II}, we briefly describe 
the generalized linear sigma model to include the broken scale symmetry,
which is used to study the medium modifications of the scalar gluon 
condensate through a scalar dilaton field.
In section \ref{III}, the leading order formula for the mass shifts of the 
quarkonium states due to the medium modification of the scalar gluon 
condensates (through the medium change of the dilaton field) 
in symmetric and asymmetric nuclear matter are discussed. The in-medium
quarkonium masses are calculated assuming the state to be a bound state
of the heavy quark and heavy antiquark using the harmonic oscillator 
potential. 
In section{IV}, we discuss the results of the present work. In subsection \ref{A}, we describe the medium modifications of the scalar fields in the chiral effective Lagrangian. The mass shifts of the charmonium and bottomonium states in dense nuclear matter are discussed in subsection \ref{B} and \ref{C}, respectively.
In section \ref{V}, we provide the summary of the consequences of the present work.
\section{Linear Sigma model WITH BROKEN SCALE INVARIANCE}
\label{II}
In this section, we describe a genralized Linear Sigma model, 
an effective Lagrangian approach 
based on the chiral $SU(2)\times SU(2)$ symmetry. The Linear Sigma model 
is genralized to incorporate 
the QCD scale invariance breaking effect through an effective potential 
in terms of the scalar dilaton
field $\chi$ as well as other fields introduced in the Lagrangian.  
The scalar dilaton field $\chi$ thus is related to the scalar gluon condensate, 
$\Big<\frac{\alpha_s}{\pi}G_{\mu\nu}^aG^{a\mu\nu}\Big>$ and the relation
is obtained by equating the trace of the energy momentum tensor
in QCD and in the generalized effective chiral model.
The Lagrangian includes an intermediate range attraction
through a scalar-isoscalar $\sigma$ meson and
scalar-isovector ${\vec\pi}$, and, a short range repulsion 
through a vector-isoscalar $\omega_{\mu}$ meson and 
a vector-isovector $\rho_{\mu}$ meson.
The Lagrangian density is given by \cite{Heide},
\bea
    \mathcal{L} = \mathcal{L}_0 - V_G + \mathcal{L'}.
    \label{c1}
\eea
The chiral symmetric, scale invariant part of the Lagrangian density 
is given by
\bea
    \mathcal{L}_0 =\frac{1}{2}\partial_{\mu}\sigma \partial^{\mu}\sigma + \frac{1}{2}\partial_{\mu}\vec{\pi}. \partial^{\mu}\vec{\pi}+ \frac{1}{2}\partial_{\mu}\chi\partial^{\mu}\chi - \frac{1}{4}\omega_{\mu\nu}\omega^{\mu\nu} + \frac{1}{2}\omega_{\mu}\omega^{\mu} G_{\omega \sigma} (\sigma^2 + \vec{\pi}^2)\nonumber\\+ \bar{\psi}_i\Big(\gamma^{\mu}\Big(i\partial_{\mu} - g_{\omega}\omega_{\mu}\Big)- g_{\sigma}\Big(\sigma + i\vec{\pi}.\vec{\tau}\gamma_5\Big) \Big)\psi_i; \quad i =p, n
    \label{c2}
    \eea
    where, the field strength tensor $\omega_{\mu\nu} = \partial_{\mu}\omega_{\nu} - \partial_{\nu}\omega_{\mu}$ corresponds to the vector meson field $\omega_{\mu}$. $\psi_i\ (i=p,n)$ are the nucleon fields in the present case of nuclear matter. Theoretically, coupling of $\omega_{\mu}$ to both of the $\sigma$ and $\chi$ fields may generate the scale-invariant $\omega$-meson mass term, $m_{\omega}$. Based on the phenomenological results of \cite{Heide}, only the coupling to $\sigma$ field is considered in our present work in the $\omega$ meson mass generation. In vacuum, the form of $m_\omega$ is, $m_\omega =G_{\omega\sigma}^{1/2}\sigma_0$, where $\sigma_0$ is the vacuum expectation value of the $\sigma$ field. The Lagrangian accounts the scale symmetry breaking of QCD (as well as the global chiral symmetry breaking) through logarithmic terms in the effective potential, $V_G (\chi, \sigma, \vec{\pi})$, which is chosen to reproduce the trace anomaly\footnote{Scale symmetry is broken by gluon condensate, in the mass less quark limit, in vacuum, which causes the non-zero trace of the energy momentum tensor in QCD} of QCD within this approach, via N\"oether's theorem \cite{Heide, heide1992},
\bea
\langle \theta_{\mu}^{\mu} \rangle = 4 V_G(\Phi_i) - \sum_i \Phi_i \frac{\partial V_G}{\partial \Phi_i} = 4 \epsilon_{vac}\Big( \frac{\chi}{\chi_0} \Big)^4.
\label{c3}
\eea
The parameter $\Phi_i$ runs over the scalar fields \{$\chi,\sigma,\vec{\pi}$\}, $\epsilon_{vac}$ is the vacuum energy, $V_G(\Phi_{i,min})$ and the $\chi_0$ is the vacuum expectation value of the scalar dilaton field $\chi$. Now by comparing the trace of the energy momentum tensor calculated in the current process (as given by eq.(\ref{c3})) with the one obtained from the general QCD Lagrangian, in the mass less quark limit, which is,
\bea
\langle\theta_{\mu}^{\mu}\rangle = \Big\langle \frac{\beta(g)}{2g} G_{\mu\nu}^a G^{\mu\nu a} \Big \rangle, 
\label{c4}
\eea
where $G^{a}_{\mu\nu}$ is the gluon field strength tensor and $\beta(g)$ is the QCD renormalization group function, considered at the one loop level, one can obtain the relation between the gluon condensate and the dilaton field as following,
\bea
 \Big\langle \frac{\b(g)}{2g} G_{\mu\nu}^a G^{\mu\nu a} \Big\rangle = 4 \epsilon_{vac}\Big( \frac{\chi}{\chi_0} \Big)^4.
 \label{c5}
\eea
The above relation shows that the gluon condensate is proportional to the fourth power of the scalar dilaton field, $\chi$. The expression of the above mentioned $\b(g)$ is,
\bea
\b(g) = -\frac{11N_c g^3}{48\pi^2}\Big(1-\frac{2N_f}{11N_c}\Big) + \mathcal{O}(g^5),
\label{c6}
\eea
where, the first part in parentheses arises from the antiscreening effect originated by the self-interaction of
the gluons and the second part, proportional to $N_f$, is the contribution of self interaction of quark
pairs i.e., screening effect \cite{carter}. The $N_c$ and the $N_f$ is the number of color and flavor quantum numbers of quarks, respectively. From Ref. \cite{heide1992, Heide}, the suggested form of the potential is given by,
  \bea
    V_G = B\chi^4 \left(ln\left(\frac{\chi}{\chi_0}\right) - \frac{1}{4}\right) - \frac{1}{2}B\delta \chi^4 ln\left({\frac{\sigma^2 + \vec{\pi}^2}{\sigma_0^2}}\right) + \frac{1}{2}B\delta\zeta^2\chi^2\Big(\sigma^2 +\vec{\pi}^2 - \frac{1}{2}\frac{\chi^2}{\zeta^2}\Big).
    \label{c7}
\eea
In eq.(\ref{c7}), the parameters are defined as, $\zeta = \chi_0/\sigma_0$, where fields with subscript 0 denote their corresponding vacuum values. The logarithmic terms contribute to the trace anomaly, such that the relation, $\epsilon_{vac} = -B\chi_0^4(1-\delta)/4$ satisfies eq.(\ref{c3}). The last term in eq.(\ref{c7}) ensures, in the vacuum, the values of the fields will take $\chi = \chi_0$, $\sigma=\sigma_0$ and ${\vec\pi} = 0$.
To maintain the physically necessary feature that $\epsilon_{vac}<0,$ for $B>0$, one needs $\delta<1$. The expression of $\delta$ is given by $\delta = 2N_f/(11N_c)$.\\
The vacuum mass of the nucleon in the Lagrangian is given by $M = g \sigma_0 = g f_{\pi}$, which is basically the Goldberger-Treiman \cite {GT} relation, taking the axial vector coupling constant $g_A$ to be 1 at this level of approximation \cite{ellis}. Therefore the vacuum value of sigma field becomes, $\sigma_0 = f_{\pi}$, where $f_{\pi}$ is the pion decay constant. The quantity $\chi_0$ is determined from the Ref. \cite{heide1992} and here it is taken as $\chi_0=\zeta \sigma_0\sim$ 148.8 MeV. 
Therefore, the relation between the gluon condensate and the dilaton field becomes, for $N_c = 3$,
\bea
\Big\langle \frac{\alpha_s}{\pi}G_{\mu\nu}^a G^{\mu\nu a}\Big\rangle = \frac{24}{33-2N_f} \Big( B(1-\delta) \chi^4 \Big);\quad \alpha_s = g^2/4\pi.
\label{c8}
\eea
In eq.(\ref{c8}), $\alpha_s$ is the QCD running coupling constant. To study the asymmetric nuclear matter, we have included vector-isovector field $\rho_{\mu}$ in the previous Lagrangian, eq.(\ref{c1}), the form of the corresponding term is \cite{Heide}, 
\bea
    \mathcal{L'} = - \frac{1}{4}{\rho}_{\mu\nu}.{ {\rho}}^{\mu\nu} + \frac{1}{2}G_{{\rho}}\chi^2{\rho}_{\mu}.{\rho}^{\mu} - \overline{\psi}\gamma^{\mu}\Big(\frac{1}{2}g_{{\rho}}\rho_{\mu}.\tau\Big)\psi,
    \label{c9}
\eea
where, the field strength tensor corresponding to $\rho_{\mu}$ meson field is $\rho_{\mu\nu} = \partial_{\mu}\rho_{\nu} - \partial_{\nu}\rho_{\mu} $ and $\psi$ is the two component column matrix of nucleons. From the 2nd term of eq.(\ref{c9}), the $\chi^2$ term turns the $\rho_{\mu}$ mass term into a scale invariant form and in the vacuum the $\rho_{\mu}$ meson mass takes the form as $m_\rho =G_{\rho}^{1/2}\chi_0$.
It is convenient to use the dimensionless ratio corresponding to the scalar fields $\chi$ and $\sigma$ as, $\phi = {\chi}/{\chi_0}$ and $\nu = {\sigma}/{\sigma_0}$, in further calculations.\\
Now, using the Lagrangian density, eq.(\ref{c1}), the coupled equations of motion for $\chi$ and $\sigma$ fields in terms of $\phi$ and $\nu$, respectively and the equations of motion for the rest of the fields i.e. for $\rho_{0}$ and $\omega_{0}$ fields are obtained under the mean field approximation,
\bea
    4B_0\phi^3(ln{\phi} - \delta ln{\nu})+ B_0 \delta\phi(\nu^2 - \phi^2)- m_{\rho}^2 \rho_0^2\phi = 0
    \label{c10}
\eea   
\bea
 B_0\delta\left(\frac{\phi^4}{\nu} - \phi^2\nu\right) + m_{\omega}^2\omega_0^2\nu - \sigma_0\sum_{i=p,n}g_{\sigma i}\rho_i^s= 0 ,
 \label{c11}
\eea
\bea
    m_{\omega}^2\nu^2 \omega_0 -\sum_{i=p,n} g_{\omega i}\rho_i = 0,
    \label{c12}
\eea
\bea
m_{\rho}^2\phi^2 \rho_0 - g_{\rho} \rho_3 = 0.
\label{c13}
\eea
 The parameter $B_0$ is defined as $B\chi_0^4$. The baryon-meson coupling constants are, from eq.(\ref{c2}), $g_{\sigma i}=g_{\sigma}$ and $g_{\omega i}=g_{\omega}$. In eqs.(\ref{c10})-(\ref{c13}), the scalar density $\rho_i^s$, number density $\rho_i$ for the $i^{th}$ baryon ($i = p, n$) and the isospin number density $\rho_3$ are defined as follows,
\bea
    \rho_i^s = \frac{\gamma_s}{(2\pi)^3}\int d^3k \frac{M_i^*}{\sqrt{k_i^2 + M_i^{*2}}},
    \label{c14}
\eea
\bea
    \rho_i = \frac{\gamma_s}{(2\pi)^3}\int d^3k,
    \label{c15}
\eea
\bea
    \rho_3 = \frac{1}{2}[{\psi}_p^{\dagger}\psi_p 
- {\psi}_n^{\dagger}\psi_n] = \frac{1}{2}[\rho_p - \rho_n] = - \eta \rho_B,
    \label{c16}
\eea
considering $\eta = {(\rho_n - \rho_p)}/(2\rho_B)$ \cite{eta} 
as the isospin asymmetry parameter and  $\gamma_s=2$ is the spin 
degeneracy factor. In the above equations, $M_i^* = M_i \nu$ 
is the effective mass of the $i^{th}$ baryon. 
The mass shifts of the heavy quarkonia through the medium 
modifications of the scalar dilaton field in dense nuclear matter 
is described in the section \ref{III}.
\section{Mass modifications of the Quarkonium states}
\label{III}
 The basic premise to study the mass modifications of the heavy quarkonium states in the nuclear medium, is demonstrated in this section. The mass modifications of the heavy quarkonium states in the medium is assumed to be due to the change in the gluon condensates, as it is illustrated in refs. \cite{DAM1, leeko, amarvdmesonTprc}. The mass shifts of the quarkonium states are expressed with the lowest dimension gluonic operators, in the QCD operator product expansion (OPE) of the correlation function of heavy quark currents.  
Thus, the mass shift of the heavy quarkonia at the leading order, in the large mass limit of heavy quarks, is dependent on the medium modification of the scalar gluon condensate  \cite{leeko, DAM1}, as given by,
\bea
    \Delta m_{\psi} = \frac{1}{18}\int d{\bf k}^2 \Big\langle \Big\vert \frac{\partial \psi (\bf k)}{\partial {\bf k}} 
\Big\vert^{2} \Big\rangle \frac{\bf k}{\frac{{\bf k}^2}{m_Q} + \epsilon}\Big(\Big<\frac{\alpha_s}{\pi} G_{\mu\nu}^a G^{\mu\nu a}\Big> - \Big<\frac{\alpha_s}{\pi} G_{\mu\nu}^a G^{\mu\nu a}\Big>_0\Big),
\label{Q1}
\eea
where,
\bea
\Big\langle \Big\vert \frac{\partial \psi (\bf k)}{\partial {\bf k}} 
\Big\vert^{2} \Big\rangle
=\frac {1}{4\pi}\int 
\Big\vert \frac{\partial \psi (\bf k)}{\partial {\bf k}} \Big\vert^{2}
d\Omega.
\label{Q2}
\eea
In the above equation, $\psi(\bf k)$ is the momentum space wave function of the corresponding quarkonium state. They are determined by solving the Schr\"odinger equation with a harmonic oscillator potential \cite{leeko,amarvdmesonTprc,amarvepja}, which will be discussed in the following section. Eq.(\ref{Q1}) involves the derivative of wave function with respect to momentum which is a measurement of color dipole size of the state. The mass shifts of the respective states are dependent on the dipole size and the binding energy of the state.
The mass of the heavy quark is denoted by $m_Q$ and taken as 1.95 GeV \cite{leeko} for charm quark and 5.36 GeV \cite{DAM1} for bottom quark to reproduce the energy splitting between the 1S and 2S states in the vacuum and the binding energy of the each respective states properly. The vacuum mass of the corresponding quarkonium state is denoted by $m_{\psi}$ and the binding energy of the heavy quark-antiquark bound state is, $\epsilon=2m_Q-m_{\psi}$. 
For $N_c = 3$ and $N_f = 2$, the form of the scalar gluon condensate (from eq.(\ref{c8})), within the chiral effective Lagrangian approach is,
\bea
    \Big\langle\frac{\alpha_s}{\pi}G_{\mu\nu}^a G^{\mu\nu a}\Big\rangle = \frac{24}{29} \Big( B(1-\delta) \chi^4 \Big).
    \label{Q3}
\eea
Then, the mass shift of the charmonium as well as bottomonium states from their respective vacuum values, are given in terms of the medium modifications of the scalar dilaton field, $\chi$ as
\bea
  \Delta m_{\psi} = \frac{4}{87} B(1-\delta)\int d{\bf k}^2 \Big\langle\Big|\frac{\partial \psi(\bf k)}{\partial {\bf k}}\Big|^2\Big\rangle \frac{\bf k}{\frac{{\bf k}^2}{m_Q} + \epsilon}\Big( \chi^4 - \chi_0^4 \Big),
  \label{Q4}
\eea
which states that the mass shift is proportional to the difference of the fourth power of the dilaton field, $\chi$ in medium and in vacuum, in the mass-less quarks limit. The mass modification of the charmonium and bottomonium states are examined using the above mass shift formula eq.(\ref{Q4}) and the produced results will be explained in the section \ref{IV}.

The wave functions of the different states of quarkonium are determined 
by treating them as a quark-antiquark state bound with a harmonic oscillator 
potential. The real space wave functions $\psi(r,\theta,\phi)$ can be obtained by solving the Schr\"odinger equation \cite{arfken},
\bea
\nabla^2\psi + \frac{2M}{\hbar^2}\Big(E - \frac{1}{2}M\omega^2 r^2\Big)\psi = 0.
\label{W1}
\eea
M is the reduced mass of heavy quarkonium system, $M = m_Q/2$, where $m_Q$ is the mass of the heavy quark, as defined before. Now using the standard separation of variables technique the complete solution of eq.(\ref{W1}) is given by \cite{amarvdmesonTprc,amarvepja, FLS},
\bea
    \psi_{N,l}(r,\theta,\phi) = A\ Y_l^m (\theta, \phi) (\beta^2 r^2)^\frac{l}{2} e^{-\frac{1}{2}\beta^2r^2} L^{(l+\frac{1}{2})}_{N-1}(\beta^2r^2),
    \label{W2}
\eea
where, $\vec{r} = \vec{r}_Q - \vec{r}_{\bar{Q}}$ \cite{voloshin}, is the relative radial coordinate and A is the normalization constant. Here $\beta$ is not the previously defined QCD $\beta$ function (as in section \ref{II}), but is defined as, $\beta^2 = M\omega/{\hbar}$ \cite{FLS}, which actually characterizes the strength of the harmonic oscillator potential of the corresponding quarkonium state. The $\beta$ parameter can be fixed \cite{beta1} by the analytic expression of r.m.s. radii of the respective quarkonium states, i.e., $<r^2>^\frac{1}{2} = (\int d^3r \psi^* r^2 \psi)^\frac{1}{2}$, and $L_a^b(x)$ is the associated Laguerre Polynomial, with the mathematical expression given below,
\bea
    L_a^b(x) = \sum_{c=0}^a (-1)^c \frac{(a+b)!}{(a-c)!(b+c)!c!}x^c.
    \label{W3}
\eea
To get the mass shift of the respective state we have to find the wave functions in the momentum space. Here we use the following property for this transformation \cite{arfken},
\bea
    \Big(f(t \textbf{r})\Big)^T(\textbf{k}) = \frac{1}{t^3} g(t^{-1}\textbf{k}),
    \label{W4}
\eea
where, $g(\textbf{k})$ is the Fourier transform of $f(\textbf{r})$. The wave functions of the quarkonium states in the momentum space are normalized as  Ref. \cite{norm} 
\bea
\int \frac{d^3k}{(2\pi)^3}\Big|\psi(\textbf{k})\Big|^2 \,=\,1.
\label{W5}
\eea
\section{Results and Discussions}
\label{IV}
The mass shifts of the charmonium
states $J/\psi$, $\psi(2S)$, $\psi(1D)$, $\chi_{c0}(1^3P_0)$, $\chi_{c1}(1^3P_1)$ and $\chi_{c2}(1^3P_2)$ with their vacuum masses (in MeV) \cite{pdg}, 3096.9, 3686.1, 3773.7, 3414.71, 3510.67 and 3556.17, respectively and the bottomonium states $\Upsilon(1S)$, $\Upsilon(2S)$, $\Upsilon_2(1D)$, $\chi_{b0}(1^3P_0)$, $\chi_{b1}(1^3P_1)$ and $\chi_{b2}(1^3P_2)$ with their vacuum masses (in MeV) \cite{pdg}, 9460.3, 10023.26, 10163.7, 9859.44, 9892.78 and 9912.21, respectively, are studied in a dense nuclear matter. The mass modifications of the corresponding states within the nuclear medium are originated from the medium modifications of the scalar dilaton field, $\chi$ in the symmetric as well as in the asymmetric nuclear medium through eq.(\ref{Q4}), for both zero and finite baryon densities up to $\rho_B=6\rho_0$ (where, $\rho_0=$ 0.153 $fm^{-3}$, is the nuclear matter saturation density \cite{eta}). In subsection \ref{A}, we discuss the in-medium behaviour of the scalar dilaton field and the scalar isoscalar field with the variation in nuclear matter density, within symmetric ($\eta = 0$) and asymmetric ($\eta = 0.5$) nuclear matter, which is to be calculated by solving the set of coupled equations written in eqs.(\ref{c10})-(\ref{c13}). Afterwards, in subsection \ref{B} and \ref{C}, we discuss the results of our present study on the mass shifts of the charmonium and bottomonium states in a dense nuclear medium.
\subsection{IN-MEDIUM BEHAVIOUR OF THE SCALAR FIELDS}
\label{A}
\begin{figure}[h!]
    \centering
    \includegraphics[width=11cm]{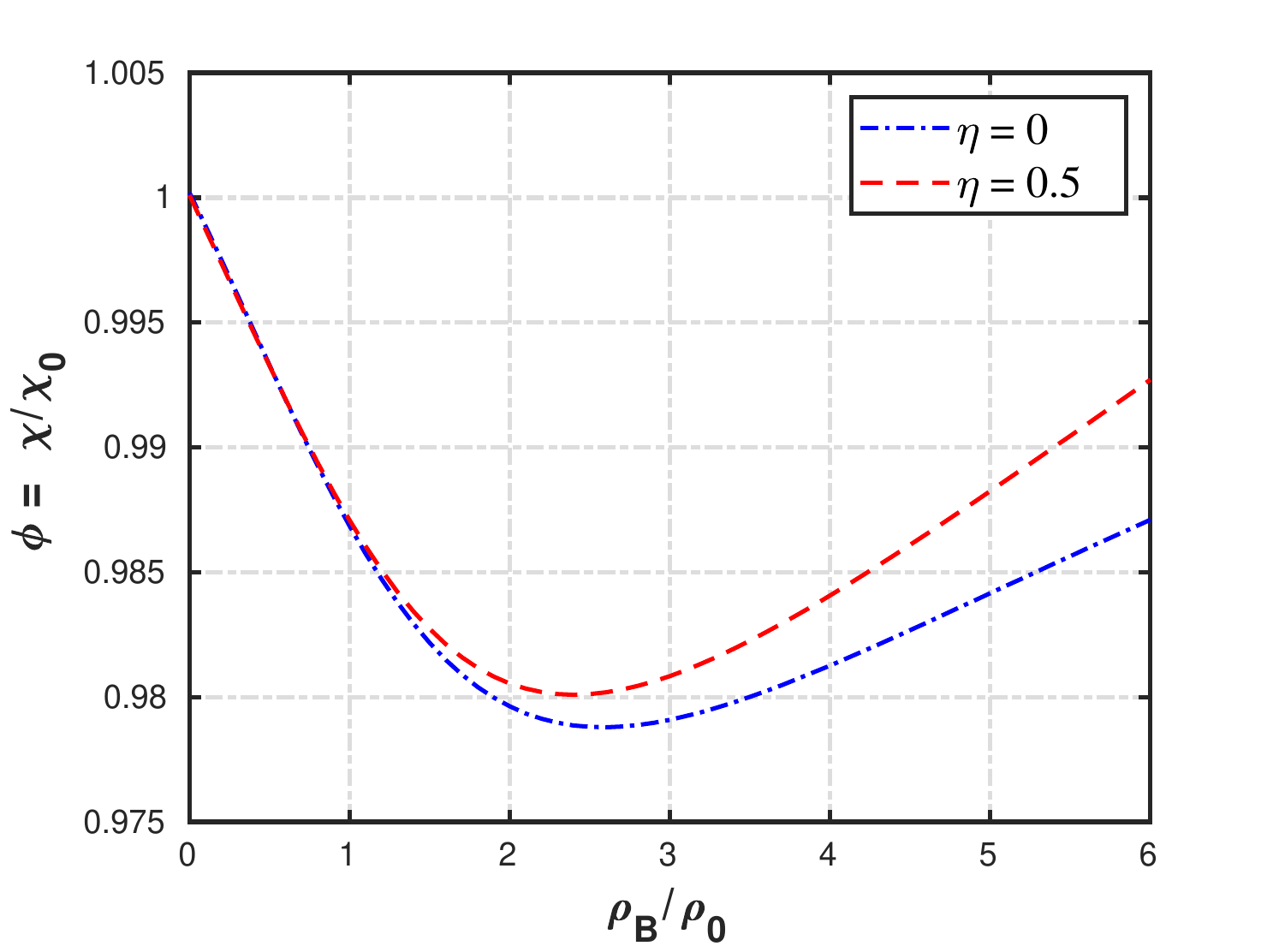}
    \caption{\raggedright{ Behaviour of the scalar field $\chi$ in terms of the ratio of $\phi=\chi/\chi_0$, is plotted as functions of the baryonic density $\rho_B$ (in units of $\rho_0$) up to $6\rho_0$, for different values of the isospin asymmetry parameter, $\eta$. We show the results for $\eta = 0.5$ (dashed line) and compare with the isospin symmetric case i.e. $\eta=0$ (dash-dotted line).}}
    \label{fig:figure1a}
\end{figure}

\begin{figure}[h!]
    \centering
    \includegraphics[width=11cm]{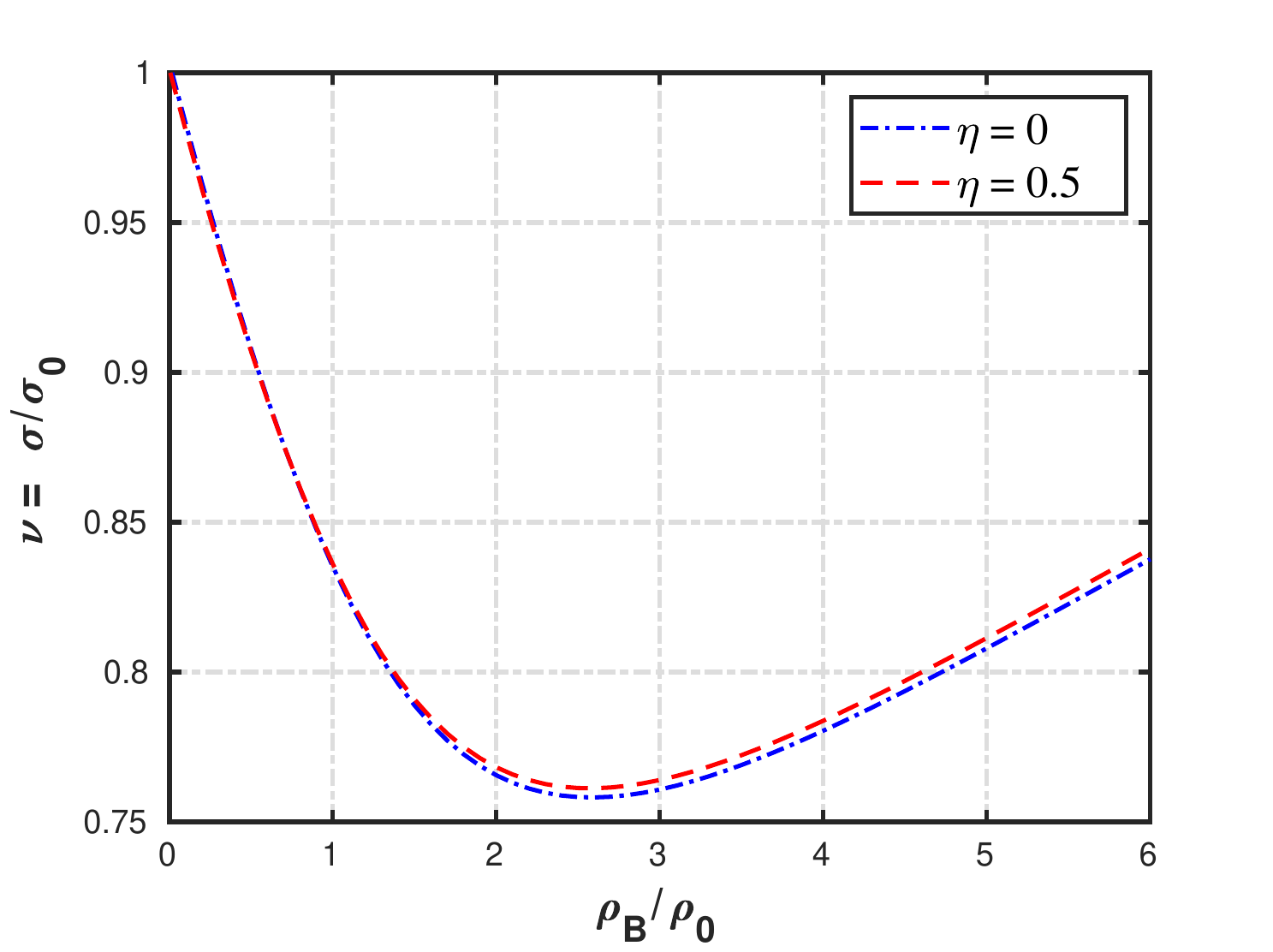}
    \caption{\raggedright{ Behaviour of the scalar field $\sigma$ in terms of the ratio of $\nu =\sigma/\sigma_0$, is plotted as functions of $\rho_B/\rho_0$ for isospin asymmetry parameter, $\eta = 0.5$ and $\eta=0$.}}
    \label{fig:figure1b}
\end{figure}

The in-medium behavior of the scalar dilaton 
field $\chi$ and the scalar isoscalar field $\sigma$
in the symmetric as well as in the asymmetric 
nuclear matter, in terms of the ratio of the field $\chi$ ($\sigma$) 
to its vacuum expectation value $\chi_0$ ($\sigma_0$), 
$\phi = \chi/\chi_0$ ($\nu = \sigma/\sigma_0$), 
are shown in Figs.\ref{fig:figure1a} (\ref{fig:figure1b}) 
as functions of the baryonic density $\rho_B$ (in units of $\rho_0$).
 A non-monotonic behaviour of $\chi$ in terms of $\phi$ and $\sigma$ in terms of $\nu$ with density, is obtained for the given values of isospin asymmetry, depicted by the predefined parameter, $\eta$. These variations of the $\phi$ and $\nu$ with baryon density and isospin asymmetry, within the effective chiral Lagrangian approach, as described in section \ref{II}, are obtained by solving the coupled equations of motion of the mesonic fields $\sigma$ (in terms of $\nu$), $\omega_{0}$, $\rho_{0}$ and $\chi$ (in terms of $\phi$), under the mean field approximation, with the parameter set given in table - \ref{tab:table1}. We already have discussed the origin of these parameters in section \ref{II}.
\vspace{0.4cm}
\begin{table}
\centering
\begin{tabular}{| c | c | c | c | c | c |}
\hline
 $\delta$ & $\big|\epsilon_{vac}\big|^{\frac{1}{4}}$(MeV) & $C_{\omega}^2$ & $C_\rho^2$ & $\sigma_0$(MeV) &$\chi_0$(MeV)\\
\hline
$\frac{4}{33}$ & 269 & 51.3 & 132 &93 & 148.8\\
\hline
\end{tabular}
\caption{ \raggedright{Parameter set used for solving the coupled equations of motion, Eq.(\ref{c10})-(\ref{c13}).}}
\label{tab:table1}
\end{table}
\vspace{-0.5cm}

The parameters $B_0$ and $C_{\omega}^2 = g_{\omega}^2 M^2/m_{\omega}^2$ are chosen to fit the nuclear matter saturation properties and the parameter $C_{\rho}^2 = g_\rho^2 M^2/m_\rho^2$ is fixed by the symmetry energy of 35 MeV \cite{Heide}. The $\delta $ parameter is suggested as 4/33 for color quantum number $N_c = 3$ and flavor quantum number $N_f = 2$, which is obtained using the one loop estimate of the $\beta_{QCD}$ function. The choice of the vacuum values of the scalar fields in this approach is already discussed in the section \ref{II}. The equations of motion of the fields [eqs.(\ref{c10})-(\ref{c13})] contain the number densities ($\rho_i;~i=p,n$) and the scalar densities ($\rho^s_i;~i=p,n$) of the nucleons, through which the effects of density, isospin asymmetry are incorporated in the scalar fields solutions.
 \begin{table}
    \centering
    \begin{tabular}{| c | c | c | c | c |}
    \hline
   \multirow{2}{*}{$\rho_B$ } & \multicolumn{2}{ c |}{$\phi = \frac{\chi}{\chi_0}$} & \multicolumn{2}{ c |}{$\nu = \frac{\sigma}{\sigma_0}$}  \\ \cline{2-5} 
   & $\eta = 0$ & $\eta = 0.5$ & $\eta = 0$ & $\eta = 0.5$  \\
    \hline
    ~~$0$~~ & ~~ 1~~ & ~~1~~ & ~~1~~ & ~~1~~\\
    \hline
    ~~${\rho_0}$~~ & ~~0.9868~~ & ~~0.9871~~ & ~~0.8351~~ & ~~0.8360~~ \\
    \hline
    ~~$2\rho_0$~~ & ~~0.9796~~& ~~0.9805~~ & ~~0.7656~~ & ~~0.7682~~ \\
    \hline
    ~~$3\rho_0$~~ & ~~0.9791~~ & ~~0.9808~~ & ~~0.7608~~ & ~~0.7640~~  \\
    \hline
    ~~$4\rho_0$~~  & ~~0.9813~~ & ~~0.9841~~ & ~~0.7805~~ & ~~0.7838~~ \\
    \hline
    ~~$5\rho_0$~~ & ~~0.9842~~& ~~0.9882~~ & 0.8080~~ & ~~0.8114~~ \\
    \hline
    ~~$6\rho_0$~~ & ~~0.9871~~ & ~~0.9927~~ & ~~0.8377~~ & ~~0.8412~~\\
    \hline
    \end{tabular}
    \vspace{0.4cm}
    \caption{\raggedright{Variation of the scalar fields in terms of the ratio $\phi$ and $\nu$, are illustrated within a density range of $\rho_B=0-6\rho_0$, for different values of the isospin asymmetry parameter, $\eta$. We show the results for $\eta = 0$ and compare with the isospin asymmetric case i.e. $\eta = 0.5$.}}
    \label{tab:table2}
\end{table} 
It is observed that the medium modifications of the scalar isoscalar field $\sigma$ (in units of $\sigma_0$) with respect to the nuclear matter density, is larger than the variation of the scalar dilaton field $\chi$ (in units of $\chi_0$). This is consistent with the relative variation of the scalar isoscalar field $\sigma$ and scalar dilaton field $\chi$ with density at zero temperature, obtained in the chiral $SU(3)$ model, which is based on the non-linear realization of chiral $SU(3)_L\times SU(3)_R$ symmetry and the broken scale invariance of QCD \cite{amarvdmesonTprc}, although the values and the gross behaviour of the fields  are different from the present approach's results. The value of $\phi$ decreases initially, and after it reaches a certain density the value starts to increase with the density which is because of the attractive and repulsive interactions considered in the Lagrangian. These fields are also sensitive to the changes of isospin asymmetry of the medium, which is incorporated in the Lagrangian through the coupling terms involving the vector isovector field $\rho_{\mu}$ with nucleons.  
For a particular density, $\phi$ ($\nu$) attains relatively higher values for a nonzero isospin asymmetry parameter $\eta$,  than the values corresponds to $\eta = 0$. With the increase of density $\rho_B$, the difference between the values of $\phi$ as well as $\nu$ at $\eta = 0$ and at $\eta=0.5$ slowly increases, i.e. at the higher densities the values of $\phi$ and $\nu$ for $\eta = 0.5$ attains much higher values than the values for $\eta = 0$, compared to the lower densities. But the difference for $\phi$ values are more prominent than the $\nu$ values. For nuclear matter saturation density $\rho_B =\rho_0$, the observed dilaton field value (in MeV) is 146.84 and 146.88 for $\eta =0$ and $\eta = 0.5$ respectively. At $\rho_B = 3\rho_0$, the values decreased to 145.69 and 145.94 MeV for $\eta = 0$ and $\eta = 0.5$ respectively. The values again start increasing and at $\rho_B = 6\rho_0$, the field value (in MeV) becomes  146.88 and 147.72 for $\eta = 0$ and $\eta = 0.5$ respectively. It can be inferred, from Figs.(\ref{fig:figure1a}, \ref{fig:figure1b}), that the medium modified values of $\chi$ and $\sigma$ are always less than their vacuum value, $\chi_0$ and $\sigma_0$, respectively which indicates the negative mass shifts of the heavy quarkonium states with density from eq.(\ref{Q4}). 
The values of $\phi$ and $\nu$ are also given for a range of density $\rho_B=0-6\rho_0$, in table - \ref{tab:table2}. Now using these values of $\chi$ and the value of $\chi_0$, the mass shifts of the desired heavy quarkonium state with density can be obtained, as it is discussed in the next subsection. 
\subsection{MASS SHIFTS OF CHARMONIUM STATES} 
\label{B}
\begin{figure}[h!]
    \centering
    \includegraphics[width=11cm]{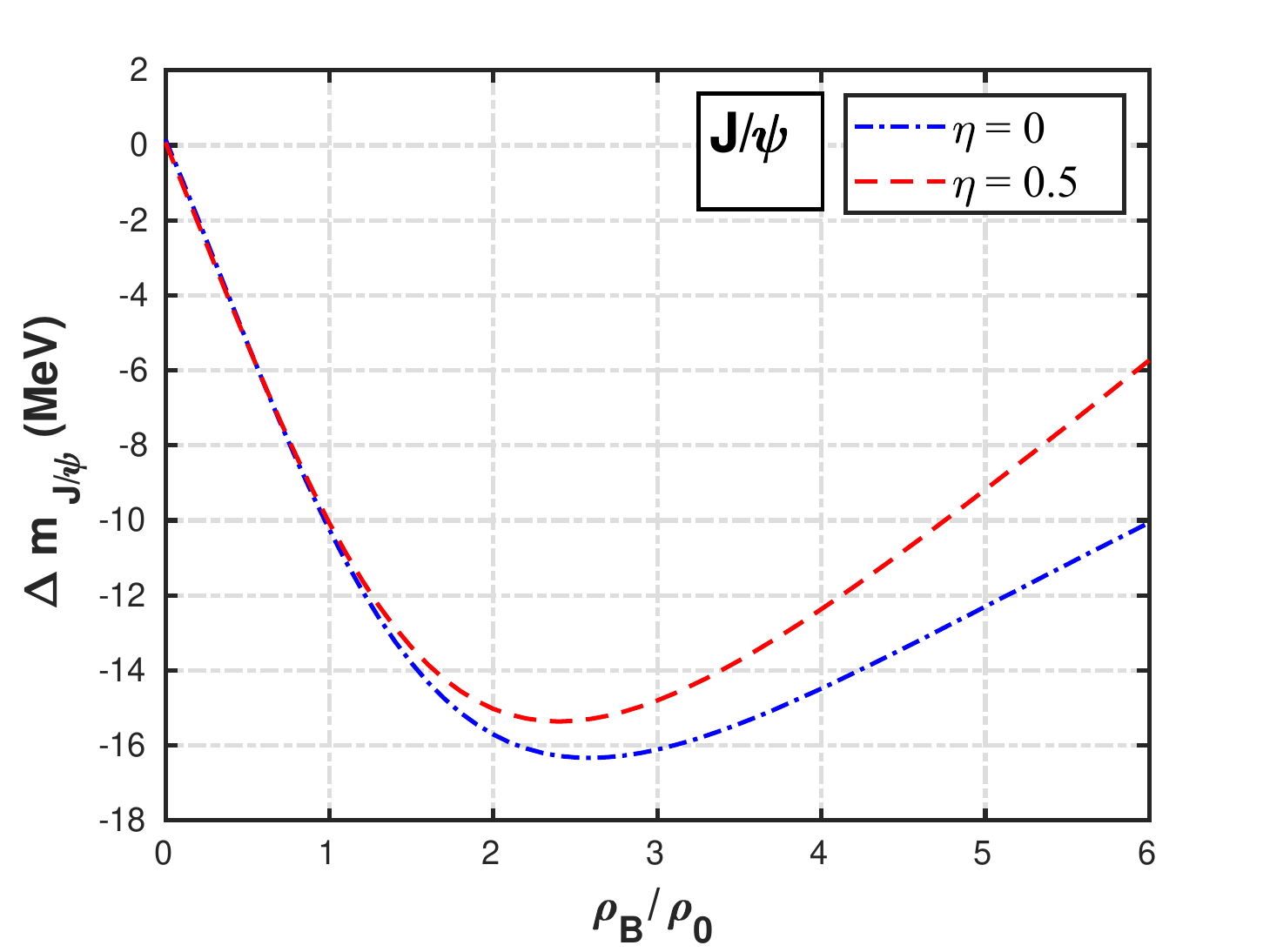}
    \caption{\raggedright{The mass shift (in MeV) of the $J/\psi$ state is plotted as a function of baryon density $\rho_B$ (in units of $\rho_0$), with isospin asymmetry parameter $\eta = 0$ and $\eta = 0.5$ .}}
    \label{fig:figure2a}
\end{figure}

\begin{figure}[h!]
    \centering
    \includegraphics[width=11cm]{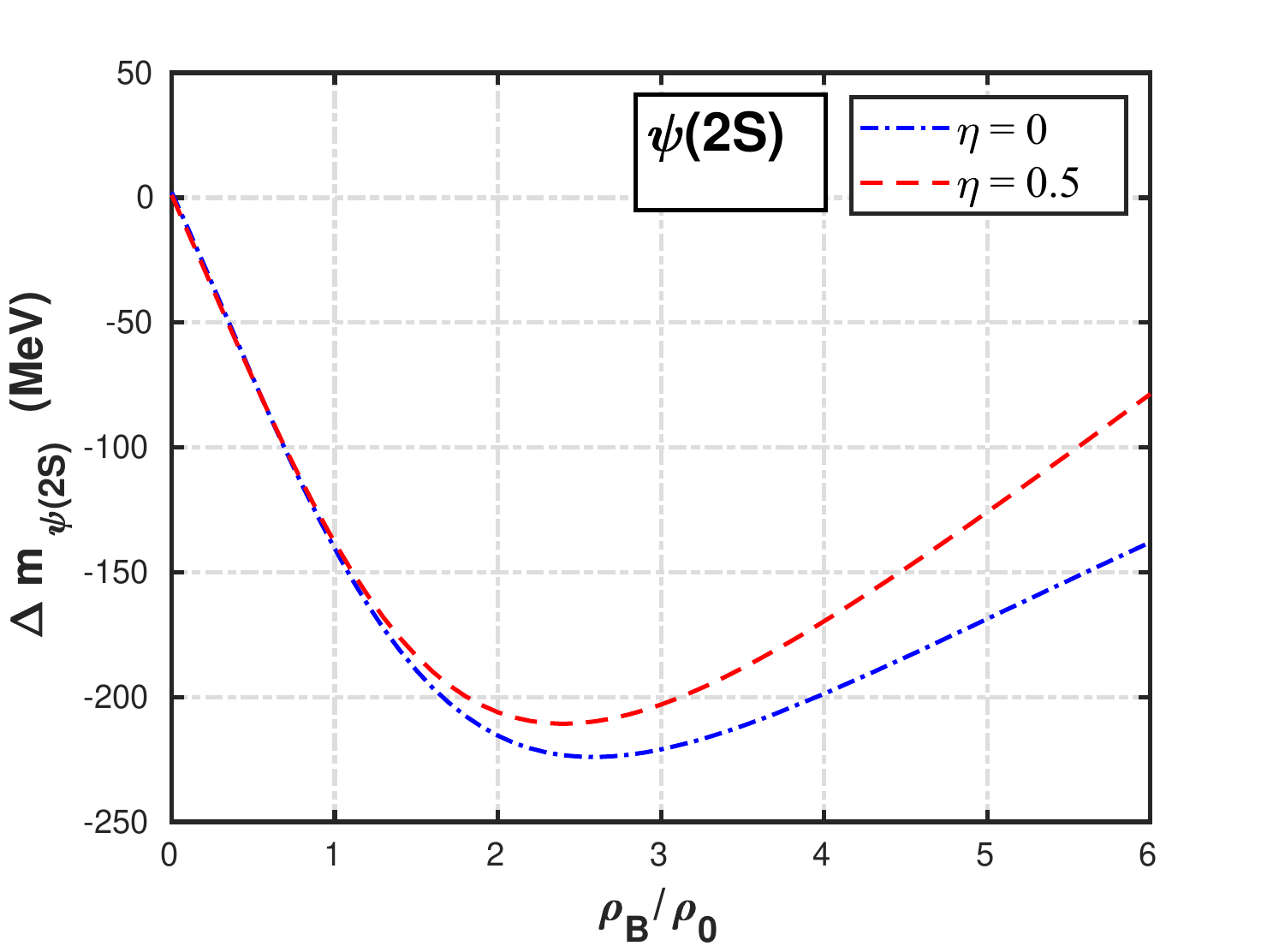}
    \caption{\raggedright{The mass shift (in MeV) of the $\psi(2S)$ state is plotted as a function of baryon density $\rho_B$ (in units of $\rho_0$), with isospin asymmetry parameter $\eta = 0$ and $\eta = 0.5$.}}
    \label{fig:figure2b}
\end{figure}

\begin{figure}[h!]
    \centering
    \includegraphics[width=11cm]{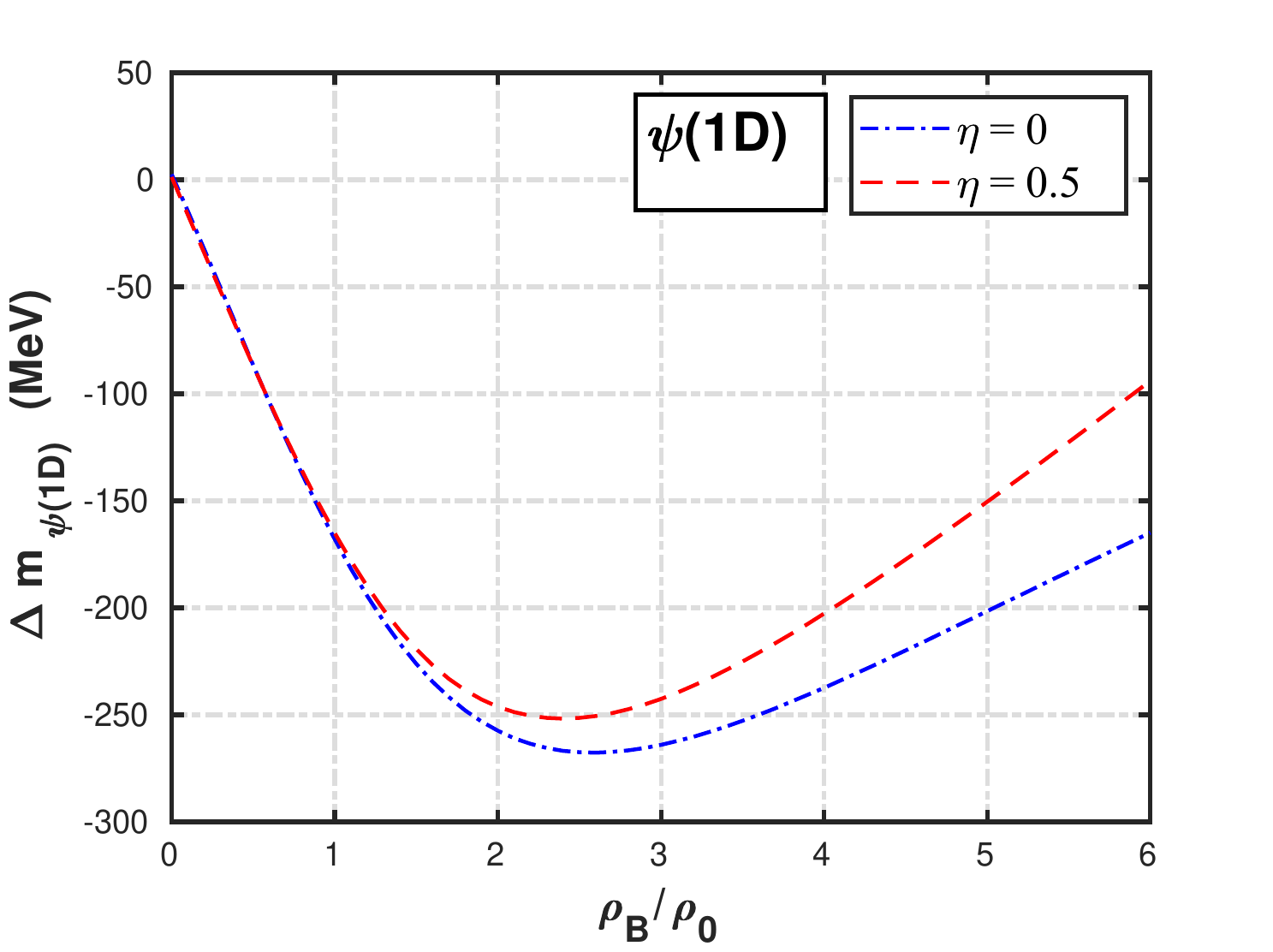}
    \caption{\raggedright{The mass shift (in MeV) of the $\psi(1D)$ state is plotted as a function of baryon density $\rho_B$ (in units of $\rho_0$), with isospin asymmetry parameter $\eta = 0$ and $\eta = 0.5$.}}
    \label{fig:figure3a}
\end{figure}

\begin{figure}[h!]
    \centering
    \includegraphics[width=11cm]{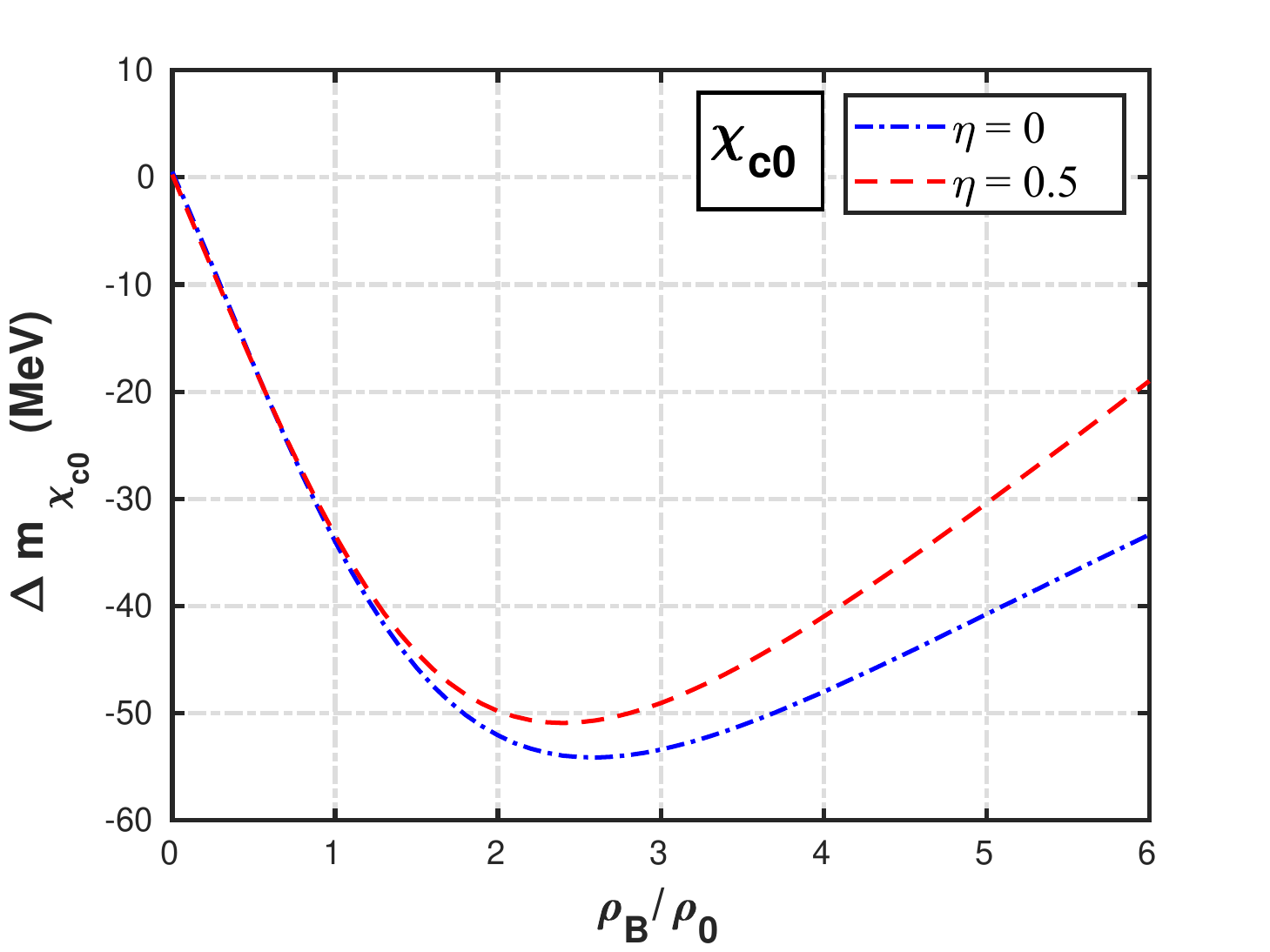}
    \caption{\raggedright{The mass shift (in MeV) of the $\chi_{c0}$ state is plotted as a function of baryon density $\rho_B$ (in units of $\rho_0$), with isospin asymmetry parameter $\eta = 0$ and $\eta = 0.5$.}}
    \label{fig:figure3b}
\end{figure}

\begin{figure}[h!]
    \centering
    \includegraphics[width=11cm]{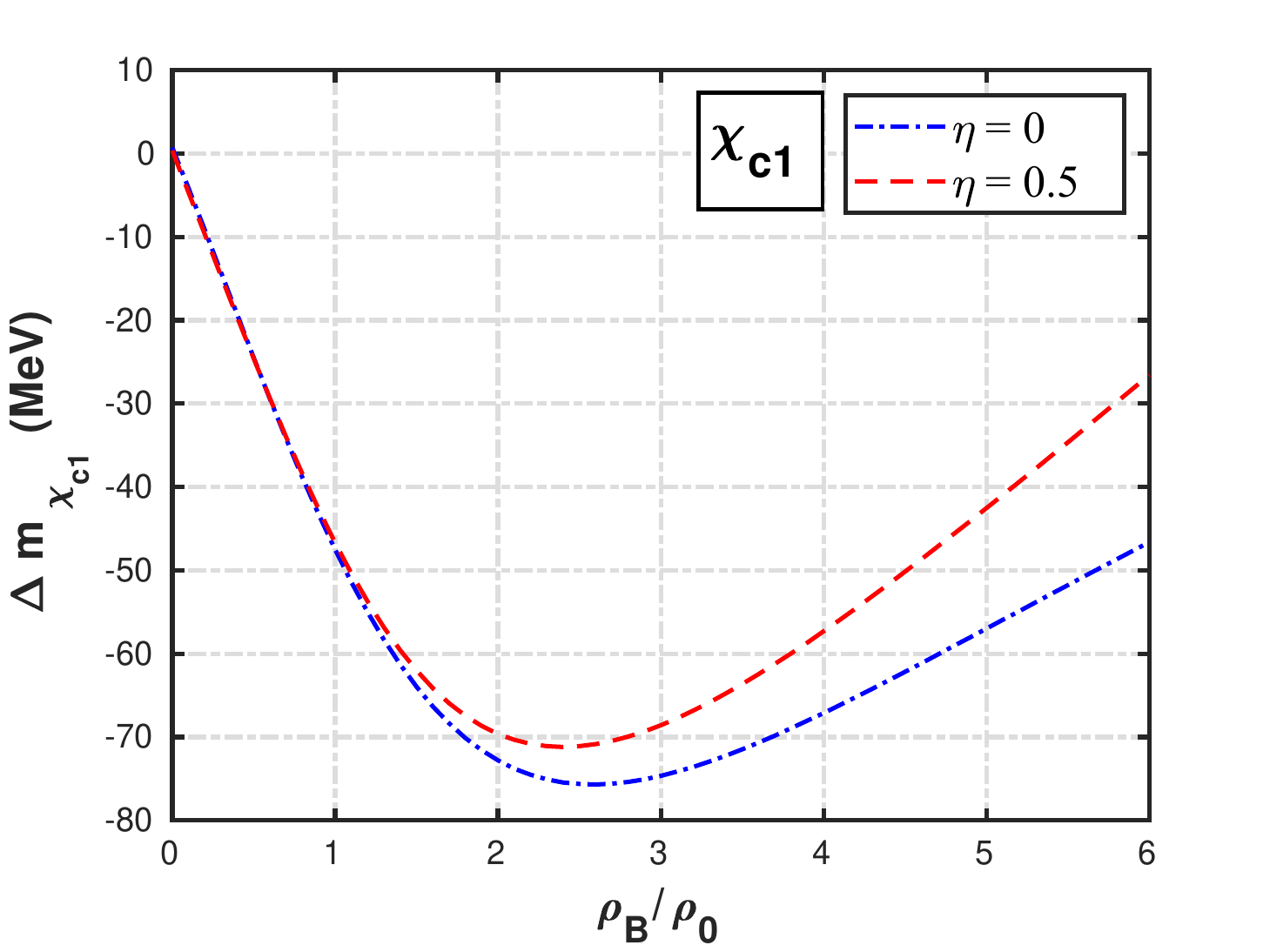}
    \caption{\raggedright{The mass shift (in MeV) of the $\chi_{c1}$ state is plotted as a function of baryon density $\rho_B$ (in units of $\rho_0$), with isospin asymmetry parameter $\eta = 0$ and $\eta = 0.5$.}}
    \label{fig:figure4a}
\end{figure}

\begin{figure}[h!]
    \centering
    \includegraphics[width=11cm]{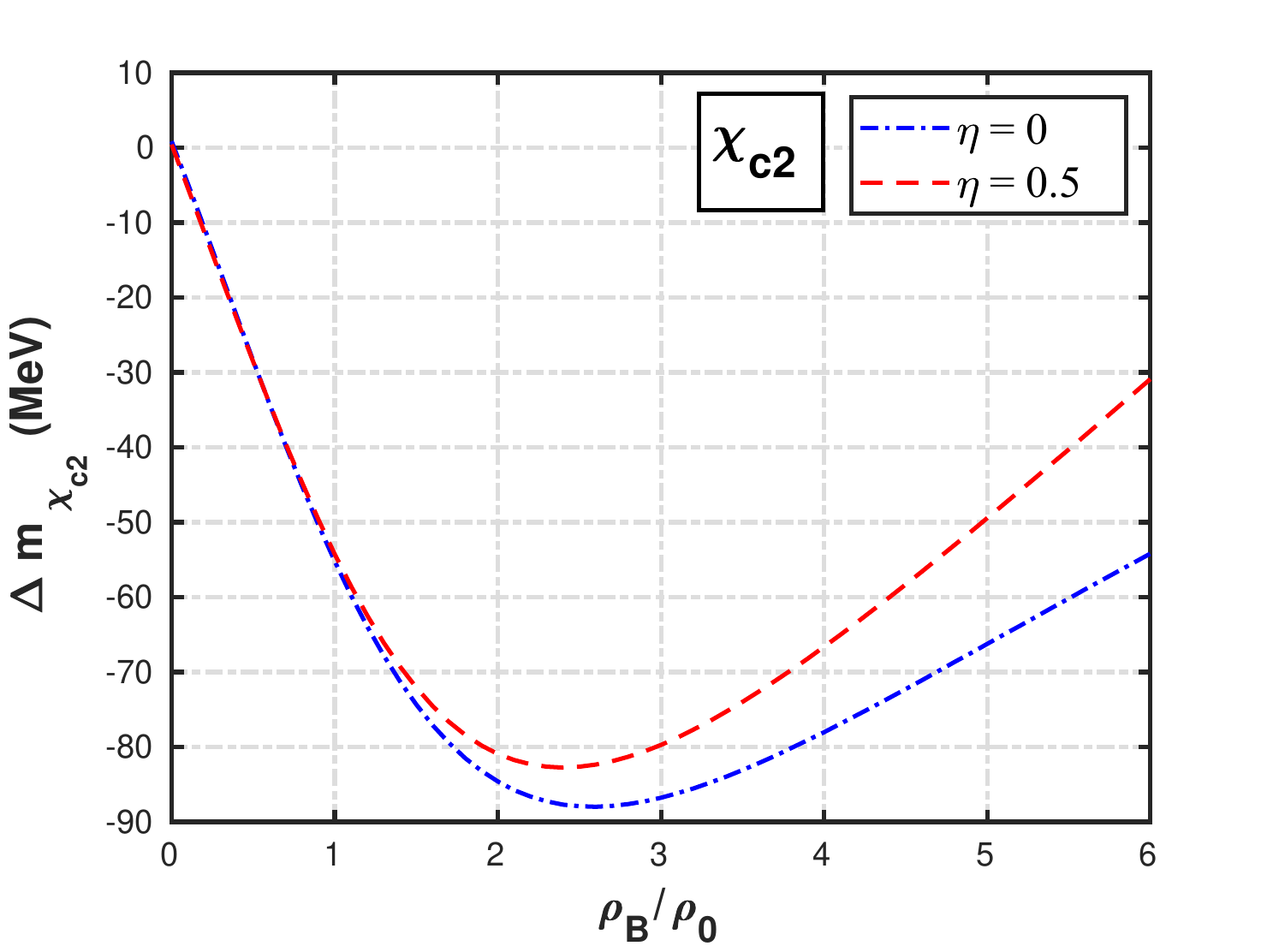}
    \caption{\raggedright{The mass shift (in MeV) of the $\chi_{c2}$ state is plotted as a function of baryon density $\rho_B$ (in units of $\rho_0$), with isospin asymmetry parameter $\eta = 0$ and $\eta = 0.5$.}}
    \label{fig:figure4b}
\end{figure}

In this subsection, we study the mass modifications of the charmonium states with the variation in nuclear matter density $\rho_B$, in the symmetric as well as asymmetric nuclear matter with the isospin asymmetry parameter, $\eta=0,\ 0.5$, respectively. The charmonium states, to be investigated in the current study, are $J/\psi$, $\psi(2S)$, $\psi(1D)$, $\chi_{c0}$, $\chi_{c1}$ and $\chi_{c2}$. Mass shifts of these states depend on the difference of the in-medium scalar gluon condensate from its value at $\rho_B=0$, in terms of the fourth power of the scalar dilaton field $\chi$, with the proportionality constant determined by the integral part of the eq.(\ref{Q4}), in terms of the wave function of the specific charmonium state. As discussed in the subsection \ref{A}, it can be seen that the in-medium values of $\chi$ are less than its vacuum value. Therefore, from eq.(\ref{Q4}), it can be stated that the masses of the various charmonium states must decrease than the vacuum masses and the respective states will get a negative mass shift in the finite densities.\\
The study shows a decrease from their vacuum values with the increase of density from $\rho_B=0$ to $6\rho_0$ in a particular pattern. The magnitude of the mass shift first increases up to around $3\rho_0$ and then decreases steadily up to $6\rho_0$, which basically follows the medium behaviour of the dilaton field. In further, the magnitude of the mass shifts from their respective vacuum values are seen to have lesser drop in the asymmetric nuclear matter, which is because of the relatively higher values of the dilaton field $\chi$ for $\eta=0.5$ than the case of $\eta =0$, as shown in Fig.\ref{fig:figure1a}. In table - \ref{tab:table3}, we have tabulated the mass shifts for each state under study, up to $6\rho_0$ in symmetric as well as asymmetric nuclear matter.
The Figs.(\ref{fig:figure2a}, \ref{fig:figure2b}, \ref{fig:figure3a}, \ref{fig:figure3b}, \ref{fig:figure4a} and \ref{fig:figure4b}) show the mass shift for $J/\psi$, $\psi(2S)$, $\psi(1D)$, $\chi_{c0}$, $\chi_{c1}$ and $\chi_{c2}$ for $\eta=0$ and 0.5, in the nuclear matter with the relative baryon density, $\rho_B/\rho_0$.
\begin{table}
    \centering
    \begin{tabular}{| c | c | c | c | c | c | c | c |}
    \hline
    ~~$\rho_B$~~ & ~~$\eta$~~ & ~~$\Delta m _{J/\psi}$~~ & ~~$\Delta m_{\psi(2S)}$~~ & ~~$\Delta m_{\psi(1D)}$~~ & ~~$\Delta m_{\chi_{c0}}$~~ & ~~$\Delta m_{\chi_{c1}}$~~ & ~~$\Delta m_{\chi_{c2}}$~~\\
    \hline
    \multirow{2}{*}{$\rho_0$} & 0.5 & -10.06 & -137.97 & -164.84 & -33.34 & -46.62 & -54.18 \\
    & 0 & -10.29 & -141.12 & -168.59 & -34.10 & -47.68 & -55.42\\
    \hline
    \multirow{2}{*}{$2\rho_0$} & 0.5 & -15.05 & -206.50 & -246.71 & -49.90 & -69.78& -81.09\\
    & 0 & -15.73 & -215.74 & -257.75 & -52.14  & -72.90 & -84.72 \\
    \hline
    \multirow{2}{*}{$3\rho_0$} & 0.5 & -14.83 & -203.42 & -243.03 & -49.16 & -68.73 & -79.88 \\
    & 0 & -16.10 & -220.86 & -263.87 & -53.37 & -74.63 & -86.73 \\
    \hline
    \multirow{2}{*}{$4\rho_0$} & 0.5 &  -12.34 & -169.29 & -202.26 & -40.91 & -57.20 & -77.86\\
    &  0 & -14.46 & -198.27 & -236.88  & -47.91  & -66.99 & -83.80 \\
    \hline
    \multirow{2}{*}{$5\rho_0$} & 0.5 & -9.22 & -126.42 & -151.03 & -30.55 & -42.72 & -49.64\\
    &  0 & -12.27 & -168.26 & -201.02  & -40.66  & -56.85 & -66.07\\
    \hline
    \multirow{2}{*}{$6\rho_0$} & 0.5 & -5.74 & -78.74 & -94.07 & -19.03 & -26.60 & -30.92 \\
    & 0 & -10.06  & -137.97 & -164.84 & -33.34  & -46.62 & -54.18\\
    \hline
    \end{tabular}
    \vspace{0.8em}
    \caption{\raggedright{The mass shifts $\Delta m$ in MeV of the charmonium states (a) $J/\psi$ (b) $\psi(2S)$ (c) $\psi(1D)$ (d) $\chi_{c0}$ (e) $\chi_{c1}$ and (f) $\chi_{c2}$ are tabulated for a density range $0-6\rho_0$ for $\eta=0.5,\ 0$.}}
    \label{tab:table3}
\end{table} 
The wave functions of the charmonium states are determined by assuming the harmonic oscillator potential for the quark-antiquark bound state, as we have discussed in section \ref{III}. Resulted wave functions are Gaussian in nature, with the values of the
strength parameter ($\b$) of the wave function, determined by the root mean squared radii of $J/\psi$,  $\chi_{c1}(1P)$, $\psi(2S)$ and $\psi(1D)$ obtained as (0.47 fm)$^2$,(0.74 fm)$^2$, (0.96 fm)$^2$ and (1 fm)$^2$ respectively \cite{eichten_2}, yielding their values as 0.52 GeV, 0.42 GeV, 0.38 GeV and 0.37 GeV \cite{amarvdmesonTprc,amarvepja, F1}. For the other 1P states, $\chi_{c0}$ and $\chi_{c2}$, we extract the values of $\b$ from a linear interpolation of the vacuum mass versus $\b$ graph \cite{betachi}, drawn for the previously determined $\b$ values. The values of $\b$ \cite{betachi} for $\chi_{c0}$ and $\chi_{c2}$ are obtained as 0.45 GeV and 0.41 GeV, respectively. From the above values, it can be interpreted that $\b$ values are larger for the  smaller size charmonium states. Due to the mass drop in the charmonium states, the strength of the harmonic oscillator wave function, $\b$ can get modified from which we can obtain an estimate for the size of the charmonium state in the nuclear medium by determining the r.m.s. radii of the corresponding state, which is already studied in the literature \cite{DAM1} for bottomonium states. This can further affect the scattering cross sections and the decay widths of those states \cite{FLS}.

This investigation shows that, at a particular density, the excited states are observed to have larger mass drop as compared to the ground state, which is because of the size of the dipole and the binding energy of the corresponding state. At saturation density, the mass shift of the $J/\psi$ state is -10.29 MeV for $\eta=0$ and -10.06 MeV for $\eta=0.5$ from its vacuum mass, compared with the mass drop of around -8 MeV \cite{leeko}, using QCD second order stark effect formalism and -8.6 MeV \cite{amarvdmesonTprc}, using the leading order mass formula with the chiral $SU(3)_L\times SU(3)_R$ model. The mass shifts (in MeV) of the higher states $\psi(2S)$ and $\psi(1D)$ are obtained as -141.12 and -168.59 in case of $\eta=0$ and -137.97 and -164.84 in case of $\eta=0.5$ at $\rho_B=\rho_0$. As compared to the mass shift values of refs. \cite{amarvdmesonTprc, amarvepja}, the obtained values of mass modifications are shifted towards the higher magnitude, in this approach.

At $\rho_B=3\rho_0$, there is higher mass drop of amount (in MeV) -16.10 (-14.83), -220.86 (-203.42) and -263.87 (-243.03) for $\eta=0~(0.5)$ for $J/\psi$, $\psi(2S)$ and $\psi(1D)$ respectively. The amount of mass drop then decreases, as is predicted by the nature of the $\chi$ field, till $6\rho_0$. As shown in table \ref{tab:table3}, the mass drop at $\rho_B=6\rho_0$, is of larger amount for $\eta=0$ as compared to the case of $\eta=0.5$. These are (in MeV) -10.06 (-5.74), -137.97 (-78.74) and -164.84 (-94.07) for $J/\psi$, $\psi(2S)$ and $\psi(1D)$ states of charmonium, respectively, at $\rho_B=6\rho_0$ and $\eta=0~(0.5)$, which shows a greater decrease of mass in symmetric nuclear matter than in asymmetric nuclear matter. This result shows that the asymmetry effect is more distinctive in the higher density region.
In this investigation, we also study the mass shift of the P wave states of charmonium, where it shows a little lesser shift compared to the higher states of charmonium, which is quite obvious. The initial drop in mass (in MeV) of the 1P states from their vacuum masses is about -34.10, -47.68 and -55.42 for $\eta=0$ and -33.34, -46.62 and -54.18 MeV for $\eta=0.5$ for $\chi_{c_0}$, $\chi_{c_1}$ and $\chi_{c_2}$ states, respectively, at $\rho_0$. In comparison to this results, the mass shifts for these states are obtained around -28, -40 and -46 MeV for $\eta=0$, within the chiral $SU(3)_L\times SU(3)_R$ model. Around $3\rho_0$, the mass shifts of these states get maximum drop, i.e., -53.37 (-49.16), -74.63 (-68.73) and -86.73 (-79.88) MeV for $\eta=0(0.5)$, in this approach. Then these mass behaviour traces the same way of the nature of the dilaton field up to $6\rho_0$. This study indicates a distinctive behaviour of the charmonium states in contrary to the results in references \cite{amarvdmesonTprc, amarvepja}, where the mass shifts follow a monotonic decrease up to $6\rho_0$. Therefore, the mass modifications can appreciably change the in-medium partial decay widths of charmonium states going to open charm mesons or, other lowest lying charmonium. This may have further observable consequences in the production of the charmonia and of open charm mesons in the heavy ion collision experiments where high density matter can be produced.
\subsection{MASS SHIFTS OF BOTTOMONIUM STATES}  
\label{C}
\begin{figure}[h!]
    \centering
    \includegraphics[width=11cm]{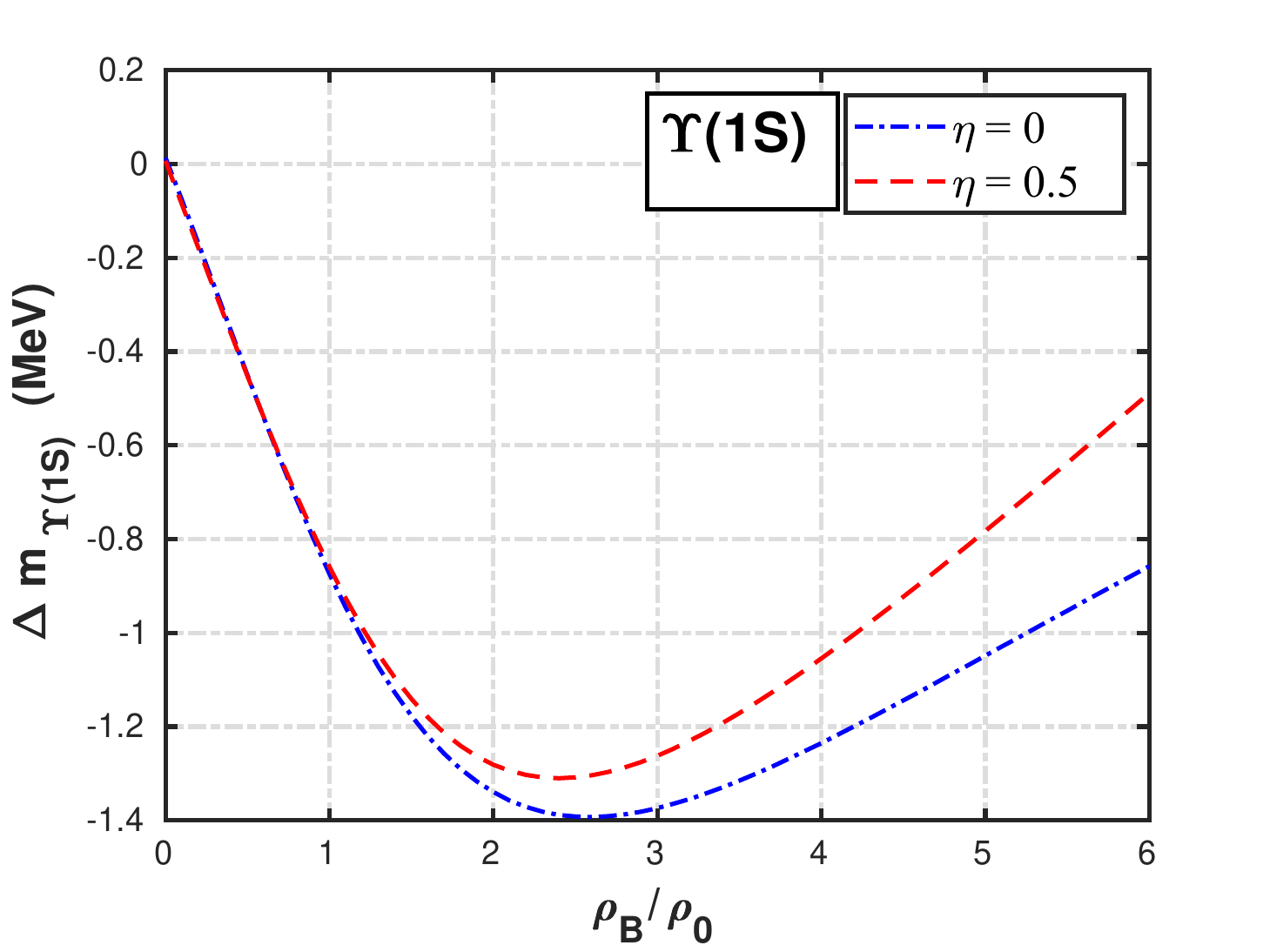}
    \caption{\raggedright{The mass shift (in MeV) of the $\Upsilon(1S)$ state is plotted as a function of baryonic density $\rho_B$ (in units of $\rho_0$) up to $6\rho_0$, with isospin asymmetry parameter $\eta = 0$ and $\eta = 0.5$.}}
    \label{fig:figure5a}
\end{figure}

\begin{figure}[h!]
    \centering
    \includegraphics[width=11cm]{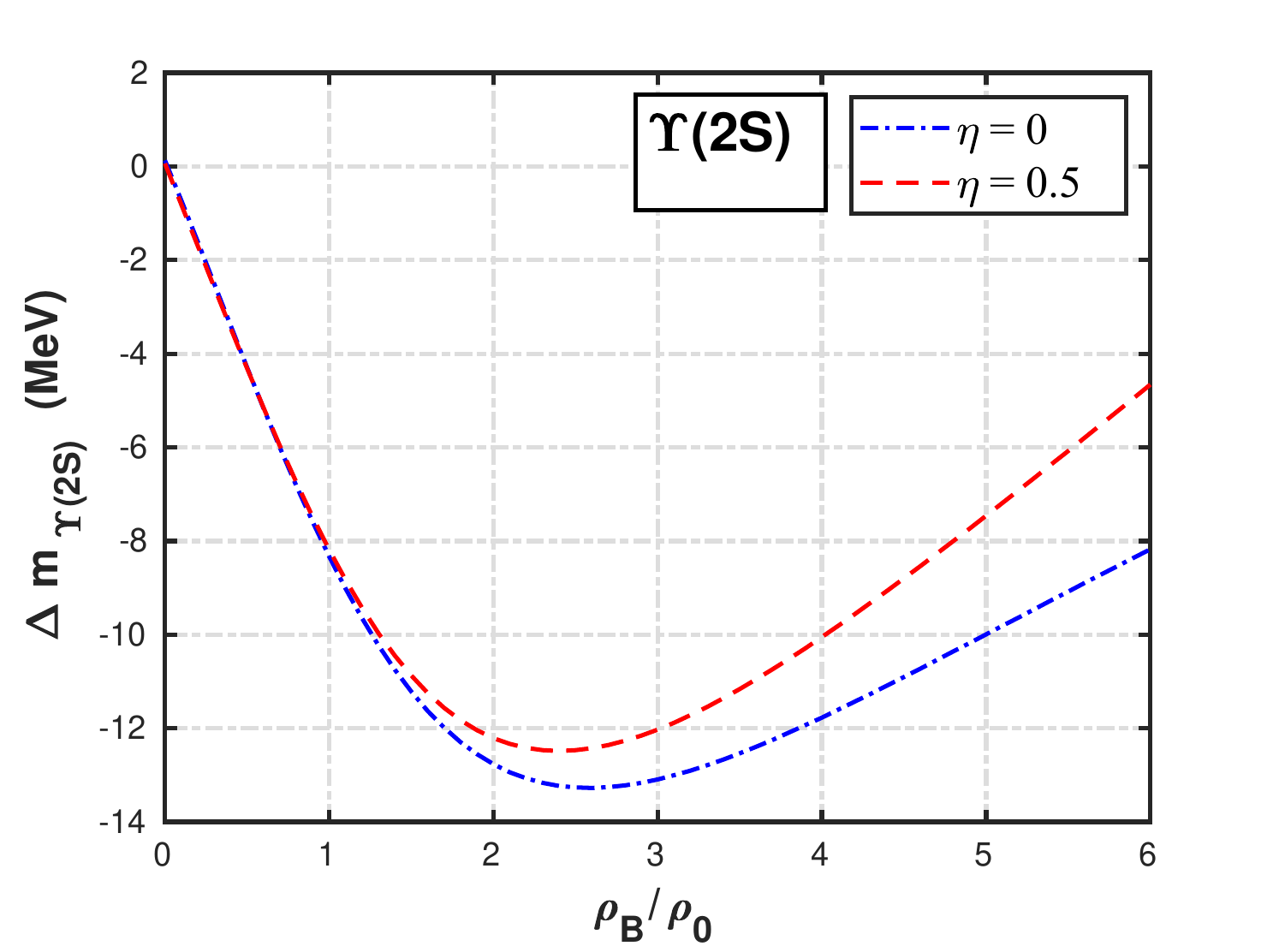}
    \caption{\raggedright{The mass shift (in MeV) of the $\Upsilon(2S)$ state is plotted as a function of baryonic density $\rho_B$ (in units of $\rho_0$), with isospin asymmetry parameter $\eta =0,\ 0.5$.}}
    \label{fig:figure5b}
\end{figure}

\begin{figure}[h!]
    \centering
    \includegraphics[width=11cm]{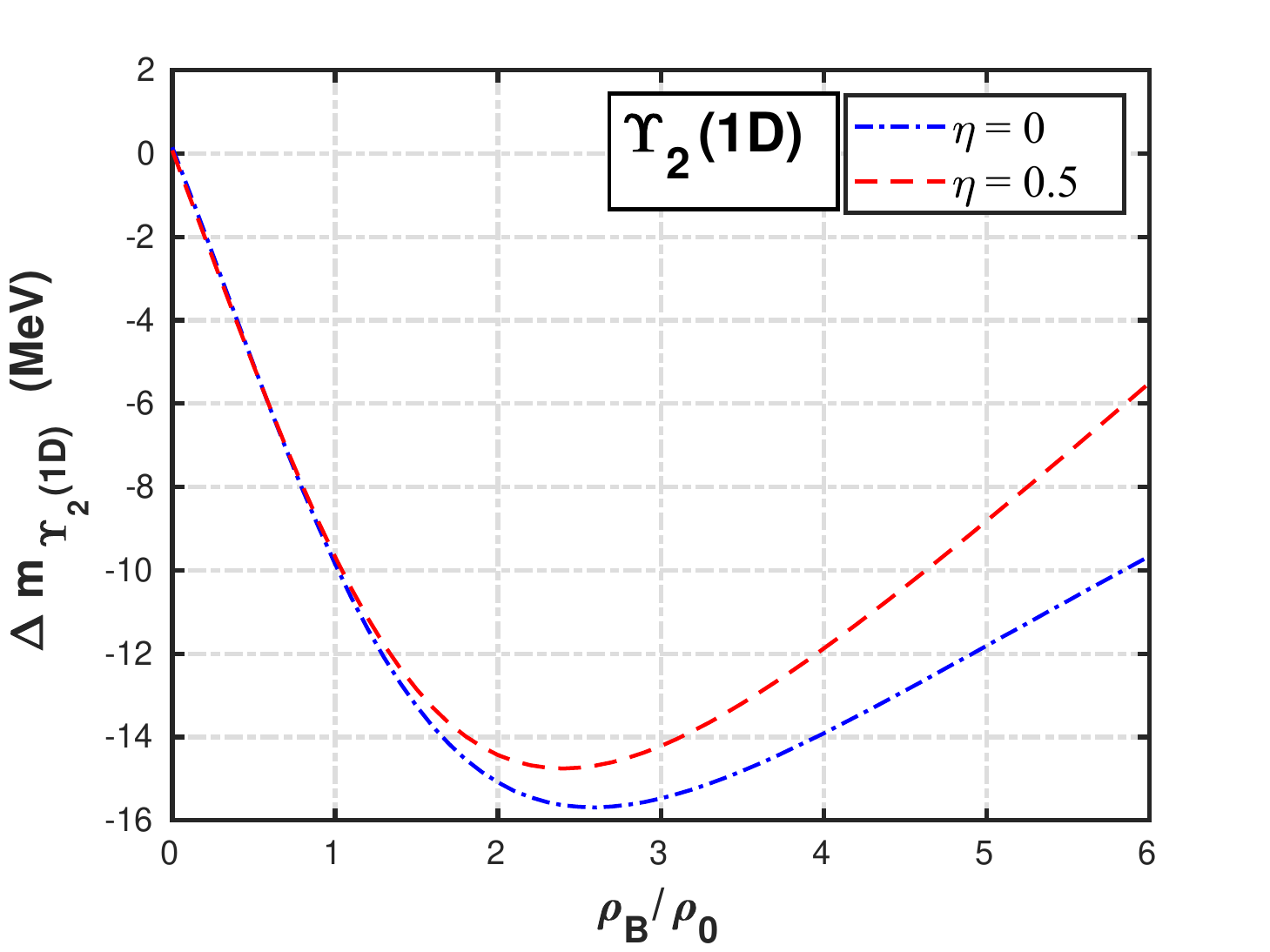}
    \caption{\raggedright{The mass shift (in MeV) of the $\Upsilon_2(1D)$ state is plotted as a function of baryonic density $\rho_B$ (in units of $\rho_0$), with isospin asymmetry parameter $\eta =0,\ 0.5$.}}
    \label{fig:figure6a}
\end{figure}

\begin{figure}[h!]
    \centering
    \includegraphics[width=11cm]{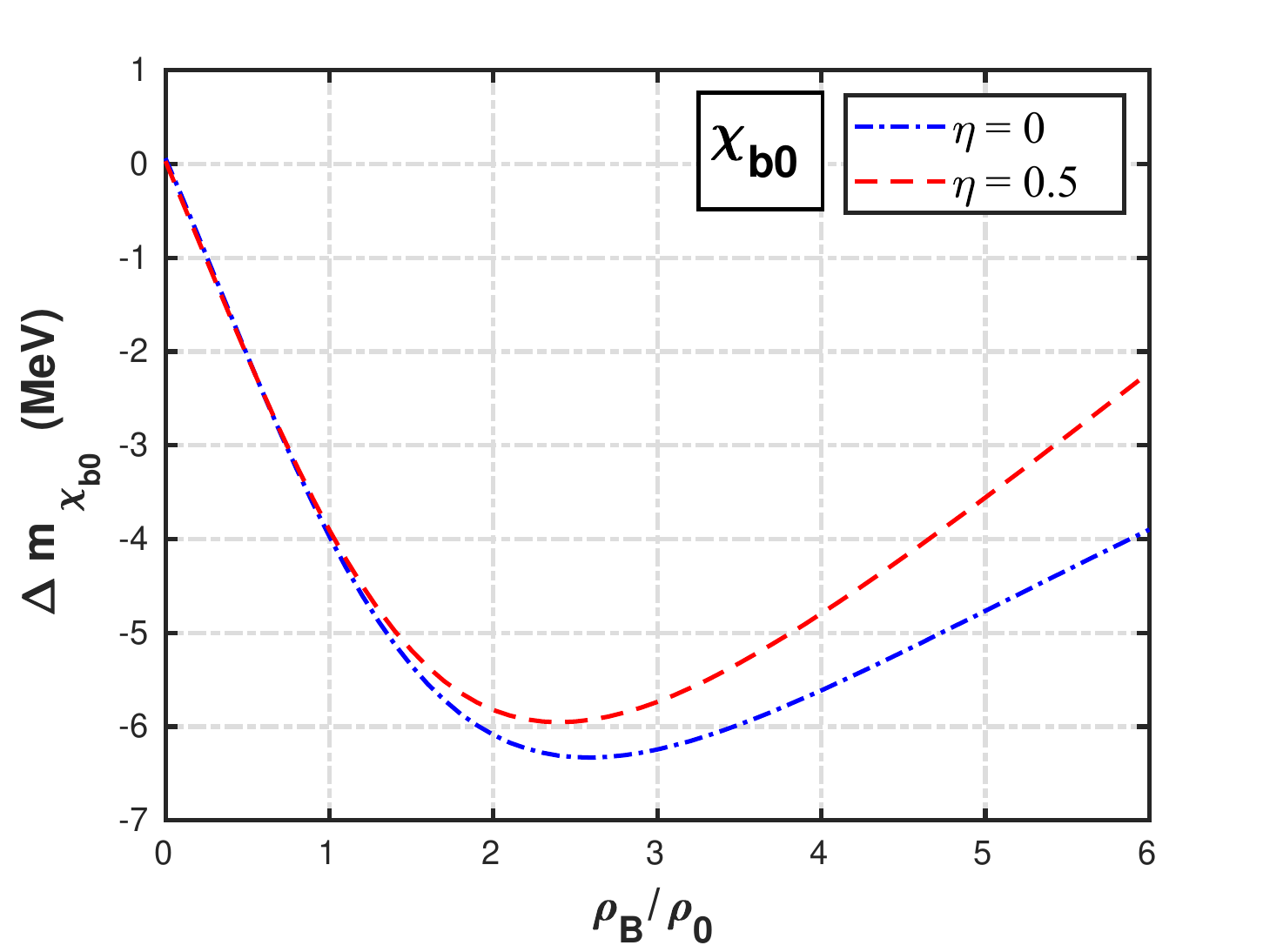}
    \caption{\raggedright{The mass shift (in MeV) of the $\chi_{b0}$ state is plotted as a function of baryonic density $\rho_B$ (in units of $\rho_0$), with isospin asymmetry parameter $\eta =0,\  0.5$ .}}
    \label{fig:figure6b}
\end{figure}

\begin{figure}[h!]
    \centering
    \includegraphics[width=11cm]{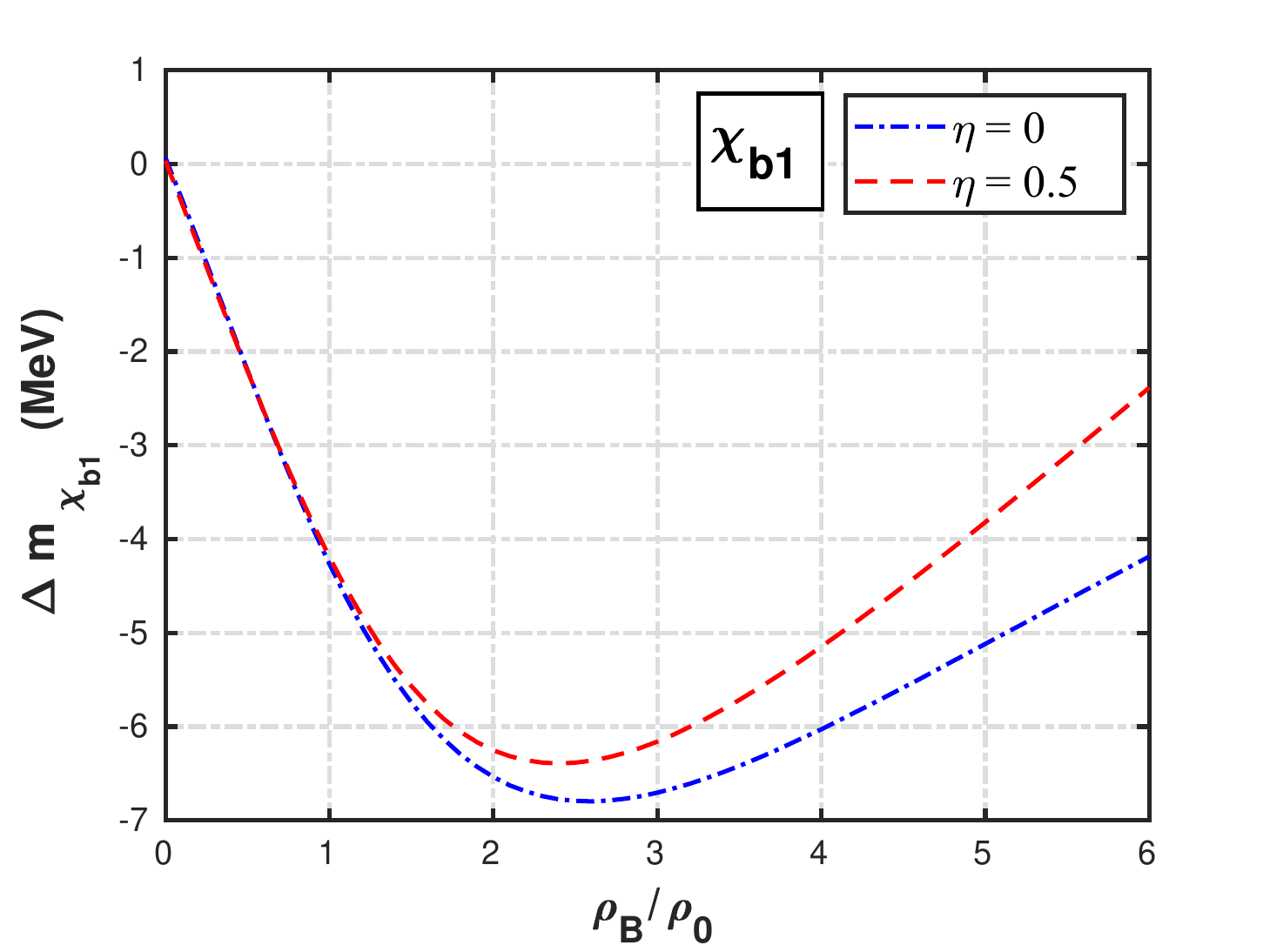}
    \caption{\raggedright{The mass shift (in MeV) of the $\chi_{b1}$ state is plotted as a function of baryonic density $\rho_B$ (in units of $\rho_0$), with isospin asymmetry parameter $\eta = 0$, $\eta = 0.5$.}}
    \label{fig:figure7a}
\end{figure}

\begin{figure}[h!]
    \centering
    \includegraphics[width=11cm]{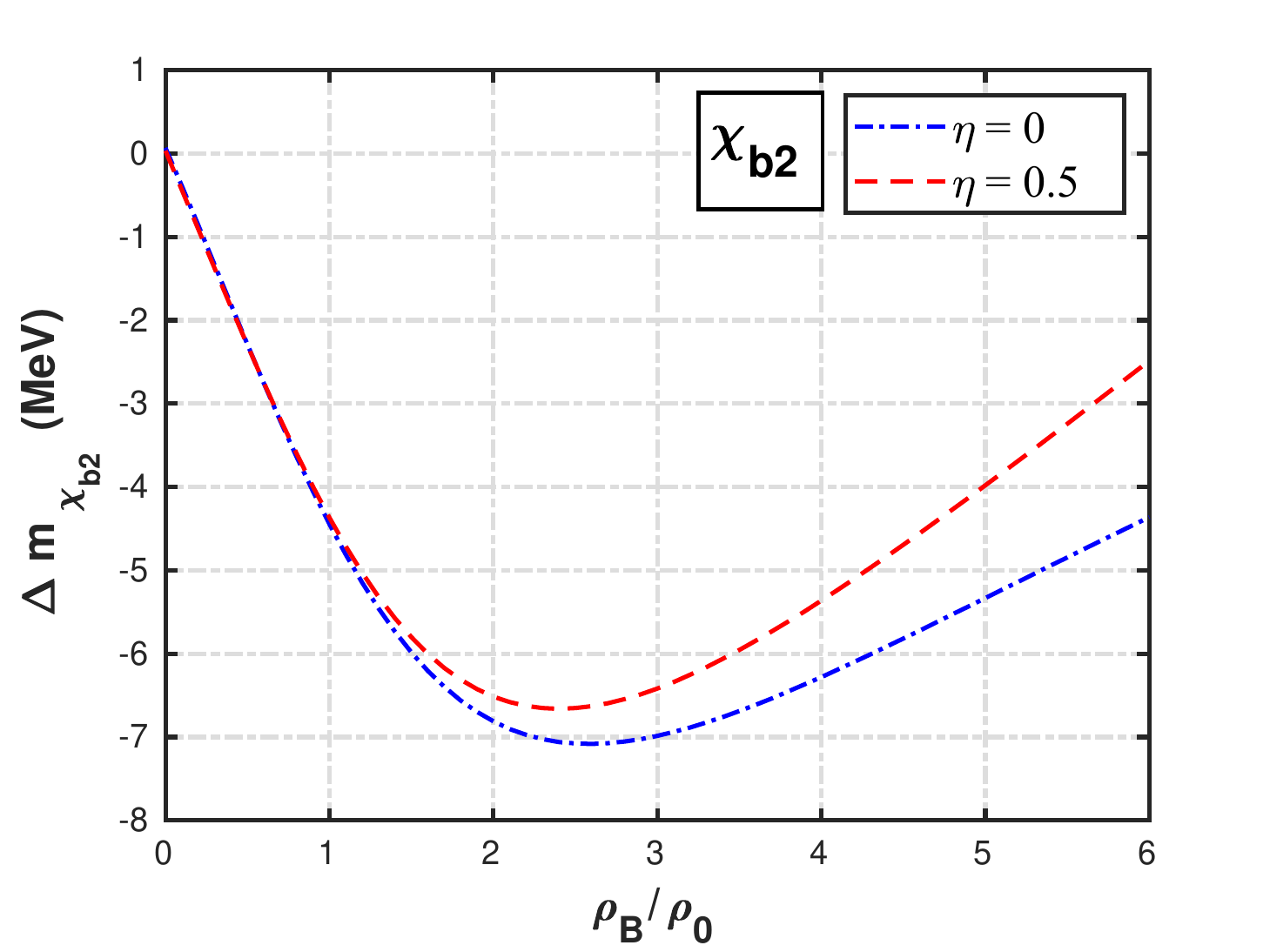}
    \caption{\raggedright{The mass shift (in MeV) of the $\chi_{b2}$ state is plotted as a function of baryonic density $\rho_B$ (in units of $\rho_0$), with isospin asymmetry parameter $\eta = 0$, $\eta = 0.5$.}}
    \label{fig:figure7b}
\end{figure}

In this subsection, we study the mass modifications of the bottomonium states at the same conditions as we have discussed before for the charmonium states. The bottomonium states under study are $\Upsilon(1S)$, $\Upsilon(2S)$, $\Upsilon_2(1D)$, $\chi_{b0}$, $\chi_{b1}$ and $\chi_{b2}$. As we have discussed for the charmonium states, the mass shifts of the bottomonium states are also proportional to the difference between the in-medium gluon condensate and its vacuum value with the proportionality constant depending on the binding energy and the dipole size of the corresponding state. Later on, this mass modification become dependent on the dilaton field and get mass drop in the finite densities. The considered density range for this work is $0-6\rho_0$. We observe a similar mass drop pattern for bottomonium states also, with a much lesser drop as compared to the charmonium states. The states experience an initial higher drop up to around $3\rho_0$ and then get a lesser drop up to $6\rho_0$. Due to the relatively higher values of the dilaton field $\chi$ for $\eta=0.5$ than the values of $\eta =0$ case, the magnitude of the mass modifications in the medium from their respective vacuum values are seen to have lesser drop in the asymmetric nuclear matter. In table \ref{tab:table4}, we have filed up the mass shifts of the bottomonium states up to $6\rho_0$ in symmetric as well as asymmetric nuclear matter.
\begin{table}
    \centering
    \begin{tabular}{| c | c | c | c | c | c | c | c |}
    \hline
    ~~$\rho_B$~~ & ~~$\eta$~~ & ~~$\Delta m _{\Upsilon(1S)}$~~ & ~~$\Delta m_{\Upsilon(2S)}$~~ & ~~$\Delta m_{\Upsilon(1D)}$~~ & ~~$\Delta m_{\chi_{b0}}$~~ & ~~$\Delta m_{\chi_{b1}}$~~ & ~~$\Delta m_{\chi_{b2}}$~~\\
    \hline
    \multirow{2}{*}{$\rho_0$} & 0.5 & -0.86 & -8.17 & -9.66 & -3.90 & -4.19 & -4.36 \\
    & 0 & -0.88 & -8.36 & -9.88 & -3.99 & -4.28 & -4.46\\
    \hline
    \multirow{2}{*}{$2\rho_0$} & 0.5 & -1.28 & -12.23 & -14.46 & -5.83 & -6.27& -6.53\\
    & 0 & -1.34 & -12.78 & -15.11 & -6.09 & -6.55 & -6.82 \\
    \hline
    \multirow{2}{*}{$3\rho_0$} & 0.5 & -1.26 & -12.05 & -14.24 & -5.75 & -6.17 & -6.43 \\
    & 0 & -1.37  & -13.09 & -15.47 & -6.24 & -6.70 & -6.98 \\
    \hline
    \multirow{2}{*}{$4\rho_0$} & 0.5 &  -1.05 & -10.03 & -11.85 & -4.78 & -5.14 & -5.35\\
    & 0 & -1.23 & -11.75 & -13.88  & -5.60 & -6.02 & -6.27 \\
    \hline
    \multirow{2}{*}{$5\rho_0$} & 0.5 & -0.78 & -7.49 & -8.85 & -3.57 & -3.83 & -3.99\\
    & 0 & -1.04 & -9.97 & -11.78 & -4.75 & -5.10 & -5.32\\
    \hline
    \multirow{2}{*}{$6\rho_0$} & 0.5 & -0.49 & -4.66 & -5.51 & -2.22 & -2.39 & -2.49 \\
    & 0 & -0.86  & -8.17 & -9.66 & -3.90 & -4.19 & -4.36\\
    \hline
    \end{tabular}
    \vspace{0.8em}
    \caption{\raggedright{The mass shifts $\Delta m$ in MeV of the bottomonium states (a) $\Upsilon(1S)$ (b) $\Upsilon(2S)$ (c) $\Upsilon_2(1D)$ (d) $\chi_{b0}$ (e) $\chi_{b1}$ and (f) $\chi_{b2}$ are tabulated for a density range $0-6\rho_0$ for $\eta=0.5,\ 0$.}}
    \label{tab:table4}
\end{table}    
Figures (\ref{fig:figure5a}, \ref{fig:figure5b}, \ref{fig:figure6a}, \ref{fig:figure6b}, \ref{fig:figure7a} and \ref{fig:figure7b}) show the mass shift for $\Upsilon(1S)$, $\Upsilon(2S)$, $\Upsilon_2(1D)$, $\chi_{b0}$, $\chi_{b1}$ and $\chi_{b2}$ with the relative baryon density, $\rho_B/\rho_0$, in the symmetric and asymmetric nuclear matter.
The wave functions of the bottomonium states are same as the wave functions of the charmonium states. The prime difference is arising from the strength parameter ($\b$) of the wave function which are determined from the root mean squared radii of $\Upsilon(1S)$, $\Upsilon(2S)$, $\Upsilon(3S)$ and $\Upsilon(4S)$ states, obtained by the decay width fitting, as (0.1843 fm)$^2$, (0.4026 fm)$^2$, (0.5925 fm)$^2$ and (0.8449 fm)$^2$ respectively \cite{eichten_2}, yielding their $\b$ values as 1.31 GeV, 0.92 GeV, 0.78 GeV and 0.64 GeV. The $\b$ values for the other states, i.e., $\chi_{b0}$, $\chi_{b1}$, $\chi_{b2}$ and $\Upsilon_2(1D)$ (using experimental observables same as the $\Upsilon(1D)$ state as per the current PDG version \cite{pdg}) can be extracted from a linear interpolation (extrapolation for the case of $\Upsilon_2(1D)$) of the vacuum mass versus $\b$ graph \cite{betachi}, drawn for the previously determined $\b$ values. The values of $\b$ for $\chi_{b0}$, $\chi_{b1}$, $\chi_{b2}$ and $\Upsilon_2(1D)$ are obtained as 0.99 GeV, 0.97 GeV, 0.96 GeV and 0.86 GeV, respectively. For the larger upsilon states the $\b$ values are smaller. Due to the mass drop in the bottomonium states, the strength of the harmonic oscillator wave function, $\b$ can be modified as discussed in the previous subsection.

In this investigation, the higher states are observed to have larger mass drop as compared to the lower state because of the respective binding energy and the dipole size of the state. The mass drops of corresponding bottomonium states in symmetric nuclear matter i.e., $\eta~=~0$ for specific density are as follows, -0.88 MeV ($\Upsilon(1S)$), -8.36 MeV ($\Upsilon(2S)$), -9.88 MeV ($\Upsilon_2(1D)$), -3.99 MeV ($\chi_{b0}(1P)$), -4.28 MeV ($\chi_{b1}(1P)$) and -4.46 MeV ($\chi_{b2}(1P)$) at $\rho_0$; -1.34 MeV ($\Upsilon(1S)$), -12.78 MeV ($\Upsilon(2S)$), -15.11 MeV ($\Upsilon_2(1D)$), -6.09 MeV ($\chi_{b0}(1P)$), -6.55 MeV ($\chi_{b1}(1P)$) and -6.82 MeV ($\chi_{b2}(1P)$) at $2\rho_0$;  -1.37 MeV ($\Upsilon(1S)$), -13.09 MeV ($\Upsilon(2S)$), -15.47 MeV ($\Upsilon_2(1D)$), -6.24 MeV ($\chi_{b0}(1P)$), -6.70 MeV ($\chi_{b1}(1P)$) and -6.98 MeV ($\chi_{b2}(1P)$) at $3\rho_0$;  -1.23 MeV ($\Upsilon(1S)$), -11.75 MeV ($\Upsilon(2S)$), -13.88 MeV ($\Upsilon_2(1D)$), -5.60 MeV ($\chi_{b0}(1P)$), -6.02 MeV ($\chi_{b1}(1P)$) and -6.27 MeV ($\chi_{b2}(1P)$) at $4\rho_0$;  -1.04 MeV ($\Upsilon(1S)$), -9.97 MeV ($\Upsilon(2S)$), -11.78 MeV ($\Upsilon_2(1D)$), -4.75 MeV ($\chi_{b0}(1P)$), -5.10 MeV ($\chi_{b1}(1P)$) and -5.32 MeV ($\chi_{b2}(1P)$) at $5\rho_0$ and  -0.86 MeV ($\Upsilon(1S)$), -8.17 MeV ($\Upsilon(2S)$), -9.66 MeV ($\Upsilon_2(1D)$), -3.90 MeV ($\chi_{b0}(1P)$), -4.19 MeV ($\chi_{b1}(1P)$) and -4.36 MeV ($\chi_{b2}(1P)$) at $6\rho_0$, respectively. 
For asymmetric nuclear matter i.e., $\eta~=~0.5$, the mass drop is lesser for that specific density like, -0.86 MeV ($\Upsilon(1S)$), -8.36 MeV ($\Upsilon(2S)$), -9.88 MeV ($\Upsilon_2(1D)$), -3.99 MeV ($\chi_{b0}(1P)$), -4.28 MeV ($\chi_{b1}(1P)$) and -4.46 MeV ($\chi_{b2}(1P)$) at $\rho_0$; -1.28 MeV ($\Upsilon(1S)$), -12.23 MeV ($\Upsilon(2S)$), -14.46 MeV ($\Upsilon_2(1D)$), -5.83 MeV ($\chi_{b0}(1P)$), -6.27 MeV ($\chi_{b1}(1P)$) and -6.53 MeV ($\chi_{b2}(1P)$) at $2\rho_0$;  -1.26 MeV ($\Upsilon(1S)$), -12.05 MeV ($\Upsilon(2S)$), -14.24 MeV ($\Upsilon_2(1D)$), -5.75 MeV ($\chi_{b0}(1P)$), -6.17 MeV ($\chi_{b1}(1P)$) and -6.43 MeV ($\chi_{b2}(1P)$) at $3\rho_0$;  -1.05 MeV ($\Upsilon(1S)$), -10.03 MeV ($\Upsilon(2S)$), -11.85 MeV ($\Upsilon_2(1D)$), -4.78 MeV ($\chi_{b0}(1P)$), -5.14 MeV ($\chi_{b1}(1P)$) and -5.35 MeV ($\chi_{b2}(1P)$) at $4\rho_0$;  -0.78 MeV ($\Upsilon(1S)$), -7.49 MeV ($\Upsilon(2S)$), -8.85 MeV ($\Upsilon_2(1D)$), -3.57 MeV ($\chi_{b0}(1P)$), -3.83 MeV ($\chi_{b1}(1P)$) and -3.99 MeV ($\chi_{b2}(1P)$) at $5\rho_0$ and  -0.49 MeV ($\Upsilon(1S)$), -4.66 MeV ($\Upsilon(2S)$), -5.51 MeV ($\Upsilon_2(1D)$), -2.22 MeV ($\chi_{b0}(1P)$), -2.39 MeV ($\chi_{b1}(1P)$) and -2.49 MeV ($\chi_{b2}(1P)$) at $6\rho_0$, respectively. 
 This study shows somewhat different behaviour of the bottomonium masses in contradiction to the results in ref. \cite{DAM1}, at higher densities, where there is a monotonic decrease in mass shift up to $6\rho_0$. Therefore, it can also affect significantly the decay widths and hence the production of the bottomonium and open bottom meson states due to the in-medium behavior of the bottomonium states. These results are relevant for the production of particles in the heavy ion collision experiments where highly dense matter can be produced.
\section{Summary}
\label{V}
The mass shifts of the charmonium states, $J/\psi$, $\psi(2S)$, $\psi(1D)$, $\chi_{c0}$, $\chi_{c1}$, $\chi_{c2}$ and of the bottomonium states $\Upsilon(1S)$, $\Upsilon(2S)$, $\Upsilon_2(1D)$, $\chi_{b0}$, $\chi_{b1}$, $\chi_{b2}$ are studied in nuclear matter from the medium modifications of the scalar gluon condensate calculated within a generalized effective chiral $SU(2)\times SU(2)$ Lagrangian 
incorporating the broken scalr invariance of QCD. The Lagrangian density is composed of a chiral as well as scale invariant part and a symmetry breaking part where the gluon condensate is generated through the scale invariance breaking logarithmic potential in terms of the scalar dilaton field $\chi$. The non zero trace of the energy momentum tensor in QCD and in the current approach, are compared to obtain the scalar gluon condensate through the scalar dilaton field. The coupled equations of motion of the scalar $\sigma,~\chi$ and the temporal component of the vector meson fields, $\rho_0,~\omega_0$ (under the mean field approximation) are solved by incorporating the medium effects of density, isospin asymmetry through the number ($\rho_i$), scalar ($\rho^s_i$) and isospin number ($\rho_3$) densities of the nucleons. The in-medium behaviour of the scalar gluon condensate obtained from that of the scalar dilaton field relative to the density, is somewhat different from other related studies using effective model based on different realizations of chiral symmetry. Moreover, this causes a distinct non monotonous behaviour of the mass shifts of the heavy quarkonium states with the increasing baryon density. The masses decrease with the increase of density, with an additional behaviour in the nature of the mass shifts, in which the mass shift initially increases with density up to around $3\rho_0$ and then start to decrease slowly towards the higher density region, which exactly replicates the behaviour of scalar dilaton field with density. In the asymmetric nuclear matter, the mass drops of the charmonium and the bottomonium states show a similar characteristics as before but with a lesser mass drop compared to the values of symmetric nuclear matter. In comparison with the charmonium states, bottomonium states undergo a much smaller drop in masses in the medium. Also, the mass shifts of the excited states, as compared to the ground state, of charmonium (bottomonium) get a higher mass drop. Overall, this study shows quite large mass drop for the above mentioned states in the medium, which can affect the decay widths of the heavy quarkonium states and therefore the production of charmonia (bottomonia) as well as of open charm (open bottom) mesons, in the relativistic heavy ion collision experiments, specifically at the future accelerator project at GSI and JINR, where it is feasible to produce highly dense matter.

\begin{section}*{Acknowledgements}
Amruta Mishra acknowledges financial support
from Department of Science and Technology (DST),
Government of India (project no. CRG/2018/002226).
\end{section}


\begin{thebibliography}{10}
\bibitem{Hosaka_Prog_Part_Nucl_Phys} 
A. Hosaka, T. Hyodo, K. Sudoh, Y. Yamaguchi, S. Yasui,
Prog. Part. Nucl. Phys. {\bf 96}, 88 (2017).
\bibitem{vacuum}
Vacuum Structure and QCD Sum Rules, edited by M. A. Shifman, 
Vol. 10 (Elsevier,1992).
\bibitem{pes2} 
G. Bhanot and M.E. Peskin, Nucl. Phys. B {\bf 156}, 391 (1979).
\bibitem{pes1} M.E. Peskin, Nucl. Phys. B {\bf 156}, 365 (1979).
\bibitem{leeko} Su Houng Lee and Che Ming Ko, 
Phys. Rev. C {\bf 67}, 038202 (2003).
\bibitem{ellis}
P. J. Ellis, E. K. Heide and S. Rudaz Phys. Lett. B {\bf 282}, 271 (1992).
\bibitem{heide1992}
E. K. Heide, S. Rudaz and P. J. Ellis, Phys. Lett. B {\bf 293}, 259 (1992).
\bibitem{Heide}
E. K. Heide, S. Rudaz and P. J. Ellis, Nucl. Phys. A {\bf 571}, 4 (1994).
\bibitem{Tsushima}
K. Tsushima and F. C. Khanna, Phys. Lett. B {\bf 552}, 138 (2003).
\bibitem{HIC}
W. Busza, K. Rajagopal and W. van der Schee,
Annurev-nucl. {\bf68}, 339-376 (2018).
\bibitem{GSI1}
http://www.gsi.de/fair/experiments/CBM/.
\bibitem{GSI}
S. Zschocke, T. Hilger, and B. Kampfer, Eur. Phys. Jour. A {\bf47}, 151 (2011).
\bibitem{frankfurt}
L. Frankfurt et al., Eur. Phys. J. A {\bf 56}, 171 (2020).
\bibitem{jlab}
http://www.jlab.org/.
\bibitem{nica}
http://www.jinr.ru/main-en/.
\bibitem{nica1}
J. Cleymans, Phys. Part. Nucl. Lett. {\bf 8}, 797 (2011).
\bibitem{nica2}
S. J. Brodsky, Eur. Phys. J. A, {\bf 52}, 220 (2016).
\bibitem{arfken}
G. B. Arfken and H. J. Weber, Mathematical Methods for Physicists, 6$^{th}$ edition, Elsevier (2005).
\bibitem{pdg}  
R.L. Workman et al. (Particle Data Group), Prog. Theor. Exp. Phys. 2022, 083C01 (2022).
\bibitem{open_heavy_flavour_qsr}
A. Hayashigaki , Phys. Lett. B {\bf 487}, 96 (2000);
T. Hilger, R. Thomas and B. K\"ampfer, Phys. Rev. C {\bf 79},
025202 (2009); T. Hilger, B. K\"ampfer and S. Leupold,
Phys. Rev. C {\bf 84}, 045202 (2011);
S. Zschocke, T. Hilger and B. K\"ampfer,
Eur. Phys. J. A {\bf 47} 151 (2011).
\bibitem{kimlee}
S. Kim, S. H. Lee, Nucl. Phys. A {\bf 679}, 517 (2001).
\bibitem{klingl} F. Klingl, S. Kim, S. H. Lee, P. Morath and
W. Weise, Phys. Rev. Lett. {\bf 82}, 3396 (1999).
\bibitem{amarvjpsi_qsr}
A. Kumar and A. Mishra, Phys. Rev. C {\bf 82}, 045207 (2010).
\bibitem{betachi}
A. Mishra, A. Jahan CS, S. Kesarwani, H. Raval, S. Kumar, J. Meena, Eur. Phys. J. A {\bf 55}, 99 (2019).
\bibitem{amarvdmesonTprc}
A. Kumar and A. Mishra, Phys. Rev. C {\bf 81}, 065204
(2010).
\bibitem{amarvepja}
A. Kumar and A. Mishra, Eur. Phys. A {\bf 47}, 164
(2011).
\bibitem{DAM1}
A. Mishra and D. Pathak,
Phys. Rev. C \textbf{90}, 025201 (2014).
\bibitem{DAM2}
D. Pathak and A. Mishra, Phys. Rev. C \textbf{91}, 045206 (2015).
\bibitem{AMC}
A. Jahan C.S., A. Mishra Chin. Phys. C {\bf 46}, 8, 083106 (2022).
\bibitem{1}
A. Mishra and A. Mazumdar, Phys. Rev. C {\bf79}, 024908 (2009).
\bibitem{eichten_1}
E. Eichten, K. Gottfried, T. Kinoshita, K.D. Lane and
T.M. Yan, Phys. Rev. D {\bf 17}, 3090 (1978).
\bibitem{eichten_2}
E. Eichten, K. Gottfried, T. Kinoshita, K.D. Lane and
T.M. Yan, Phys. Rev. D {\bf 21}, 203 (1980).
\bibitem{Klumberg_Satz_Charmonium_prod_review}
L. Kluberg and H. Satz, 
in Relativistic Heavy Ion Physics,
edited by R. Stock, Landolt-B\"ornstein - Group I Elementary Particles,
Nuclei and Atoms Vol. 23 (Springer, Berlin, Heidelberg,2010).
\bibitem{Quarkonia_QGP_Mocsy_IJMPA28_2013_review}
A. Mocsy, P. Petreczky and M. Strickland,
Int. Jour. Mod. Phys. A {\bf 28}, 1340012 (2013).
\bibitem{repko}
S.F. Radford and W W. Repko, Phys. Rev D {\bf 75}, 074031 (2007).
\bibitem{QMC_Krein_Thomas_Tsushima_Prog_Part_Nucl_Phys_2018}
 G. Krein, A.W. Thomas, K. Tsushima,
Prog. Part. Nucl. Phys. {\bf 100}, 161 (2018).
\bibitem{QMC1}
K. Tsushima, D. H. Lu, A. W. Thomas, K. Saito, and R. H. Landau, Phys. Rev. C {\bf59}, 2824 (1999).
\bibitem{QMC2}
J. J. Cobos-Martinez, T. sushima, G. Krein, A. W. Thomas, Phys. Lett. B {\bf 771}, 113-118 (2017).
\bibitem{F1}
A. Mishra, S.P. Mishra and W. Greiner, Int. J. Mod. Phys. E {\bf24}, 155053 (2015).
\bibitem{F2}
A. Mishra and S.P. Mishra, Phys. Rev. C {\bf 95}, 065206 (2017).
\bibitem{tolos_heavy_mesons}
R. Molina, D. Gamermann, E. Oset, and L. Tolos,
Eur. Phys. J A {\bf 42}, 31 (2009);
L. Tolos, R. Molina, D. Gamermann, and E. Oset,
Nucl. Phys. A {\bf 827} 249c (2009).
\bibitem{2}
 P. Papazoglou, D. Zschiesche, S. Schramm, J. Schaffner-Bielich, H. St\"ocker, and W. Greiner,
Phys. Rev. C {\bf59}, 411 (1999).
\bibitem{3}
D. Zschiesche, A. Mishra, S. Schramm, H. St\"ocker and W. Greiner, Phys. Rev. C {\bf70}, 045202
(2004).
\bibitem{APS}
A. Kumar, P. Parui, S. De and A. Mishra, Phys. Rev. C \textbf{100}, 015207 (2019).
\bibitem{4}
A. Mishra, A. Kumar, S. Sanyal, V. Dexheimer, S. Schramm, Eur. Phys.
J {\bf45}, 169 (2010).
\bibitem{carter}
G. Carter, P. J. Ellis, S. Rudaz, Nucl. Phys. A {\bf 603}, 367-386 (1996). 
\bibitem{GT}
T. P. Cheng, and L. F. Li, Oxford University Press (1984).
\bibitem{eta}
N. K. Glendenning, Compact stars - Nuclear Physics, Particle Physics and General Relativity, 2$^{nd}$ edition (Springer, 2000).
\bibitem{FLS}
B. Friman, S. H. Lee, T. Song, Phys. Lett. B {\bf 548}, 3-4 (2002).
\bibitem{voloshin}
M. B. Voloshin - Nuclear Physics B {\bf 154}, 365 (1979).
\bibitem{beta1}
S. H. Lee and C. M. Ko, Prog. Theor. Phys. Suppl. {\bf149},
173 (2003).
\bibitem{norm}
S. H. Lee and K. Morita, Phys. Rev. D {\bf 79}, 011501 (2009).




\end{thebibliography}
\end{document}